\pdfoutput=1
\documentclass[a4paper,12pt]{article}

\usepackage{amssymb,lscape,psfrag}
\usepackage{epsfig,color}
\usepackage{array}
\usepackage{enumerate}
\usepackage{amsmath}
\numberwithin{equation}{section}
\usepackage{graphics}
\usepackage{epsfig}
\usepackage{verbatim}
\usepackage{cancel}
\usepackage{longtable}
\usepackage{multirow}
\usepackage{float}
\usepackage[normalem]{ulem} 
\usepackage[toc,page]{appendix}
\usepackage{rotating}
\usepackage{amsmath}
\usepackage{url}
\usepackage{bbm}    
\usepackage{amssymb}
\usepackage{amsmath}
\usepackage{graphicx}
\usepackage{hyperref}
\usepackage{natbib}
\usepackage{mathabx}
\usepackage{subfig}
\usepackage{scalerel}
\usepackage{xcolor}
\hypersetup{
	colorlinks,
	linkcolor={purple!80!black},
	citecolor={blue!50!black},
	urlcolor={black!100!black}
}
\usepackage{units}

\newcommand{\cites}[1]{\citeauthor*{#1}'s \citeyearpar{#1}}

\newcommand {\vg}{\ell}

\newcommand {\dn}[1] { {\boldsymbol#1} }

\newcommand {\mt}[1] {\mathrm #1}
\newcommand {\inispace}{\renewcommand{\baselinestretch}{1.}\normalsize}

\newcommand {\be} {\begin {equation} }
\newcommand {\ee} {\end {equation} }
\newcommand {\bmath} {\begin {displaymath} }
\newcommand {\emath} {\end {displaymath} }

\newcommand {\by} {\mathbf{y}}

\newcommand {\nstar} { n^* }

\newcommand{\betab}{\boldsymbol{\beta}_\vg}
\newcommand{\betahatstar}{\widehat{\boldsymbol{\beta}}_\vg^*}

\newcommand{\invd}{1/\delta}

\newcommand{\I}{\mathbf{I}_n}
\newcommand{\X}{\mathbf{X}}
\newcommand{\XG}{\mathbf{X}_{\vg}}
\newcommand{\XT}{\mathbf{X}_{\vg}^{T}}
\newcommand{\XO}{\mathbf{X}_{0}}
\newcommand{\HL}{\mathbf{H}_{\vg}}
\newcommand{\invLamda}{\mathbf{\Lambda}_{\vg}^{-1}}
\newcommand{\Lamda}{\mathbf{\Lambda}_{\vg}}
\newcommand{\invLamdaO}{\mathbf{\Lambda}_{0}^{-1}}
\newcommand{\invLamdaOCR}{\big[\mathbf{\Lambda}_{0}^{_{(\mathrm{CR})}}\big]^{-1}}
\newcommand{\LamdaOCR}{\mathbf{\Lambda}_{0}^{_{({\mathrm{CR}})}}}

\newcommand{\LamdaO}{\mathbf{\Lambda}_{0}}
\newcommand{\IotaN}{\mathbf{1}_n}
\newcommand{\V}{\mathbf{V}_{\ell}}

\newcommand{\VCR}{\mathbf{V}_{\ell}^{_{\mathrm{(CR)}}}}
\newcommand{\invVCR}{\mathbf{V}_{\betab}^{\mathrm{(CR)}-1}}
\newcommand{\SG}{\boldsymbol{\widetilde{\Sigma}}_{\vg}}
\newcommand{\U}{\mathrm{U}}
\newcommand{\N}{\mathrm{N}}
\newcommand{\CR}{\mathrm{CR-PEP}}
\newcommand{\DR}{\mathrm{DR-PEP}}

\newcommand{\DRP}{\mathrm{DR-PEP}}

\begin{document}

\inispace 
\title{Variations of Power-Expected-Posterior Priors in Normal Regression Models}
\date{}
\author{
	Dimitris Fouskakis$^1$, Ioannis Ntzoufras$^2$ and Konstantinos Perrakis$^3$\\   
	{\it \small $^1$Department of Mathematics,
		National Technical University of Athens,} \\{\it \small Zografou Campus, Athens 
		Greece.} \\ [0.1cm]
	{\it \small $^2$Computational and Bayesian Statistics Lab, Department of Statistics,} \\{\it\small Athens University of Economics and Business, Greece.} \\ [0.1cm]
	{\it\small$^3$Department of Mathematical Sciences, Durham University, UK.}
}

\maketitle

	\begin{abstract}
		The power-expected-posterior (PEP) prior is an objective prior for Gaussian linear models, which leads to consistent model selection inference, under the M-closed scenario, and tends to favor parsimonious models. Recently, two new forms of the PEP prior were proposed which generalize its applicability to a wider range of models. The properties of these two PEP variants within the context of the normal linear model are examined thoroughly, focusing on the prior dispersion and on the consistency of the induced model selection procedure. Results show that both PEP variants have larger variances than the unit-information $g$-prior and that they are M-closed consistent as the limiting behavior of the corresponding marginal likelihoods matches that of the BIC. The consistency under the M-open case, using three different model misspecification scenarios is further investigated. 
	\end{abstract}


\section{Introduction}

\subsection{Prelude}
In this paper focus is given on the Bayesian variable selection problem in normal linear regression models.
For every model $M_\ell \in {\mathcal M}$ (the set of all models under consideration, using different combinations of the available explanatory variables) 
the sampling  distribution  $f_\ell(\cdot| \dn{ \beta }_\ell, \sigma^2,  \X_\ell)$ is specified by
\begin{equation} \label{new2-1}
( \dn{ Y } | \X_\ell, \dn{ \beta }_\ell, \sigma^2, M_\ell ) \sim
N_n ( \X_\ell \, \dn{ \beta }_\ell \, , \sigma^2 \, \I )
\, ,
\end{equation}
where $\dn{ Y } = ( Y_1, \dots, Y_n )$ is a vector containing the
responses for all subjects, $\X_\ell$ is an $n \times
d_\ell$ design matrix containing the values of the explanatory variables in
its columns ($d_\ell=p_\ell+1$; i.e. $p_\ell$ covariates plus the intercept), $\I$ is the $n \times n$ identity matrix, $\dn{ \beta
}_\ell$ is a vector of length $d_\ell$ summarizing the effects of the
covariates in model $M_\ell$ on the response $\dn{ Y }$ and $\sigma^2$
is the common error variance for all models $M_\ell$.
Finally, by $p$ we denote the total number of the explanatory variables under consideration. 

The Bayesian approach requires, for each model $M_\ell \in {\mathcal M}$, specification of prior densities $\pi_{\ell}(\dn{ \theta }_\ell)$ for the unknown model specific parameters $\dn{ \theta }_\ell = (\dn{ \beta}_\ell, \sigma^2)$  and also specification of prior model probabilities $\pi( M_\ell )$. Then using the observed data $\by = (y_1,\ldots,y_n)$, focus is usually given on the posterior probability of each model $M_\ell\in {\cal M}$, defined as
\begin{equation} \label{modelprop}
\pi ( M_\ell | \by ) = \frac { m_\ell(\by|\X_\ell) \pi ( M_\ell ) }{ \sum_{ M_{ k}
		\in {\cal M} }  m_k(\by|\X_k) \pi ( M_{ k } ) } = \left(
\sum_{ M_{ k } \in {\cal M} } PO_{  k  ,  \ell  } \right)^{ -1 } =
\left[ \sum_{ M_{ k } \in {\cal M} } BF_{  k  , \ell  } \frac{ \pi ( M_{
		k } ) }{ \pi ( M_\ell ) } \right]^{ -1 } ,
\end{equation}
where $\smash[b]{PO_{  k, \ell  } = \frac{ \pi ( M_k | \by ) }{ \pi ( M_{ \ell
		} | \by ) }}$ \vspace*{0.05in} is the \textit{posterior model odds} and $\smash[b]{BF_{
	k, \ell } = \frac{ m_k(\by|\X_k) }{ m_\ell(\by|\X_\ell) }}$ is the \textit{Bayes factor}, for comparing any two models $M_k$ and $M_{
	\ell }$ from $\cal M$. 
In (\ref{modelprop}) the quantity $m_\ell(\by|\X_\ell)$ is called the \textit{marginal likelihood} (or \textit{prior predictive distribution})	of model $M_\ell$ and is given by
$$
m_\ell(\by|\X_\ell) = \int f_{\vg}( \by | \dn{ \theta }_\ell,  \X_\ell )\pi_\ell ( \dn{ \theta }_\ell ) d\dn{ \theta }_\ell~.
$$

Regarding the prior on the model space, \cite{scott_berger_2010} argue that prior model probabilities should take into consideration  multiplicity issues inherent  in model comparisons.
When applied
to variable selection problems, this principle can be implemented by assuming that, conditionally on a random probability of inclusion  $\omega$, each predictor can enter a model independently, so that
$
\pi(M_\ell | \omega) =  
\omega^{p_\ell} (1-\omega)^{n-p_\ell}.
$
Next, a hyper-prior is assigned to $\omega$; in particular if
$
\omega \sim Beta( a_\omega, b_\omega ),
$
the resulting prior becomes
\begin{equation}
\pi(M_\ell) = \frac{B(a_\omega+p_\ell, b_\omega+p-p_\ell)}{B(a_\omega, b_\omega)},
\label{marginal_beta_binom}
\end{equation}
which is commonly known as the beta-binomial prior on model space.
The default choice $a_\omega=b_\omega=1$ results in a uniform distribution for $\omega$. Under this specification, \eqref{marginal_beta_binom} reduces to
\begin{equation}
\pi(M_\ell) = \frac{1}{p+1} \binom{p}{p_\ell}^{-1},
\label{marginal_unif_binom}
\end{equation}
which 
induces a uniform prior on model size: 
$$
\pi\big( \{ M_\ell \in {\cal M}: p_\ell=d \} \big)=1/(p+1) \mbox{~for~} d=0,1,\dots,p.
$$

Regarding the prior on model-specific regression parameters, most of the times we are \textit{a-priori} uncertain about the
validity of any competing model. This justifies the need for an objective model-selection approach in which vague prior
information is assumed, which in turn motivates the use of either diffuse-proper priors or ‘‘default" improper priors.
However, this immediately leads to further well understood difficulties. Speciffically, under proper prior distributions with
large variances (diffuse priors), the resulting Bayes factors can be highly sensitive to the chosen prior variances \citep{bartlett_57}. On the other hand, the use of default improper priors is also problematic, since such priors are defined only up
to a constant multiple and, therefore, the Bayes factor is itself a multiple of this arbitrary constant. A variety of methods
have been suggested for overcoming this problem; for a review see for example \cite{review} and \cite{forte_gonzalo_steel}.

One of the main approaches used to construct prior distributions for objective Bayes methods is the concept 
of {\it imaginary observations}. 
The basic idea, whose origin can be traced back to the work of \cite{good}
is to consider a thought experiment with an appropriate dataset that will be used to specify the normalizing constants 
involved in the Bayes factors when using improper (\textit{baseline}) priors \citep{Spiegelhalter_Smith_1982}. As we discuss in detail next, \cite{perez_berger_2002} defined the expected-posterior-prior using this idea, which motivated subsequently  \cite{fouskakis_etal_2015} to introduce the power-expected-posterior prior.

\subsection{Motivation} 
\label{mot}
\cite{perez_berger_2002} developed priors for objective Bayesian model comparison,
through utilization of the device of ``imaginary training samples". 
The \textit{expected-posterior prior} (EPP) for the parameter under a
	model is an expectation of the posterior distribution given
	imaginary observations $\by^*$ of size $n^*$. The expectation
	is taken with respect to a suitable probability measure  of a \textit{reference} model $M_0$, while the posterior distribution is computed \textit{via}
	Bayes's theorem starting from a default \textit{baseline} prior, which is typically improper.
One of the advantages of using EPPs is that impropriety
of baseline priors causes no indeterminacy in the computation of Bayes factors. However, EPPs crucially depend on the size
of the training sample. A consequence of that in variable selection problems, is that imaginary design matrices $\mt{X}^*$ (under
each competing model) must be included in the analysis, with the resulting prior (under each model) further depending
upon this choice; for a detailed discussion on this issue see
\cite{fouskakis_etal_2015}. The selection of a \textit{minimal
	training sample}, of size $n^*$, has been proposed (see for
example
\cite{berger_pericchi_2004}), to make the information
content of the
prior as small as possible, which is an appealing idea.
However, even under this setup the resulting prior can be
influential when sample size $n$ is not much
larger than the total number of parameters under the full
model; see \cite{fouskakis_etal_2015}.

The \textit{power-expected-posterior} (PEP) prior, introduced by \cite{fouskakis_etal_2015},
is an objective prior which amalgamates ideas from the power-prior \citep{ibrahim_chen_2000},
the expected-posterior prior \citep{perez_berger_2002} and the
unit-information-prior approach of \cite{kass_wasserman_95}
to simultaneously (a) produce a minimally-informative prior and (b) diminish the effect of training
samples under the EPP methodology. For a quick overview of the method, under normal linear models, see Section \ref{sec1_intro}.
The main idea is to substitute the likelihood by a \textit{density-normalized version of a power-likelihood} 
in EPP; see e.g.  
\cite{fouskakis_etal_2015,Fouskakis_Ntzoufras_2016_jcgs}{}. 

A limitation of the original PEP formulation is that the normalization of the power-likelihood does not always lead to distributions of known form, e.g. if the data come from a binomial or a Poisson distribution. \cite{Fouskakis_etal_2018_glm} tackled this problem by introducing two alternative versions of the PEP prior (named CR-PEP and DR-PER; see Section \ref{variants} for more details under normal linear models). To understand what entails each of the two alternative versions of PEP, \cite{Fouskakis_etal_2018_glm} used a simple intuitive example, where the parameter of an exponential distribution was tested using the reference prior as baseline. The original PEP prior, under this example, has the undesired property of getting more informative as the sample size grows; this is not the case under the two alternative PEP definitions.

Although the original PEP methodology under the normal linear model is well defined and studied, the properties of the two new PEP variations remain unexplored in this case. 
That is due to the fact that these priors were defined under the broad generalized linear model (GLM) framework which made the theoretical study of their properties hard because of well-known intractabilities.
Therefore, a thorough study under the normal linear setting is necessary in order to validate the new PEP versions and understand their relation to other standard methods.  
Specifically, we examine from a theoretical perspective the following:
\begin{itemize} 
\item the relationship of the new PEP versions with the original approach; 
\item model consistency under correct model specification;
\item the prior information-volume and its effect on parsimony and sparsity.   
\end{itemize}

Model selection consistency is typically investigated under the assumption of correct model specification, the so-called \textit{M-closed} setting. In general, the consistency property of proposed model selection procedures under model-misspecification has not been subjected to as much scrutiny. This latter setting is commonly referred to as \textit{M-open}; see  \cite{Rossell_Rubio2018} for a relevant recent work on this topic. Although the M-open case is not our initial motivation for this work, we further take into account this setting and investigate the robustness of the proposed PEP procedures (and of other popular methods) in simulations under model-misspecification.

The remainder of the paper is structured as follows. In Section \ref{sec1} we sketch out the formulation of the PEP prior starting from its original form and then proceeding to the two new variants. Section \ref{new_pep_normal_comparisons} focuses on certain properties of the PEP variants for the normal linear model; specifically: (i)  prior dispersion, and (ii) model selection consistency in the M-closed setting. 			        
In Section \ref{sec:sim} we present results from simulation studies guided by two overarching themes; firstly, the M-closed vs. M-open cases, and secondly, the setting where model dimensionality grows with sample size. We conclude with a discussion in Section \ref{disc}.  

\section{Background Information}
\label{sec1}

\subsection{Power Expected Posterior Priors} 
\label{sec1_intro}

\cite{fouskakis_etal_2015} and \cite{Fouskakis_Ntzoufras_2016_jcgs} studied in detail the PEP priors under the variable selection
problem in Gaussian regression models. In the former paper a PEP prior is introduced for both the model-specific
regression coefficients and the error variance, while in the latter paper another version of PEP is studied, named PCEP,
where we have a conditional prior for the regression coefficients given the error variance, which is treated as a common
nuisance parameter. Here we focus on the later case, where all posterior quantities can be derived analytically.

Specifically, we denote by $\pi_{\ell}^{\mathrm{N}}(\dn{\beta}_\ell\,, \, \sigma^2) = \pi_{\ell}^{\mathrm{N}}(\dn{\beta}_\ell\,|\sigma^2)\pi^N(\sigma^2)$ the baseline prior of the parameters of model  $M_{\ell}$. We assume that in $\mathcal M$
there exists a model $M_0$,  with parameters $\dn{\beta}_0$ and $\sigma^2$, sampling distribution $f_0(\cdot| \dn{\beta}_0, \sigma^2)$ and baseline prior 
$\pi_{0}^{\mathrm{N}}(\dn{\beta}_0\,, \, \sigma^2) = \pi_{0}^{\mathrm{N}}(\dn{\beta}_0\,|\sigma^2)\pi^N(\sigma^2)$, 
which is nested into each of the remaining models and we consider it as a reference model. This is the typical case in the variable selection problem, studied in this paper. 
Given then a set of imaginary data $\by^* = (y^*_1,\dots,y^*_{n^{*} \color{black}})^T$ and a positive power-parameter $\delta$, that is used to regulate, essentially,  the contribution of the imaginary data on the ``final" prior, we introduce the density-normalized power-likelihood, under model $M_\ell$, given by
\begin{equation}
f_{\vg}( \by^* | \betab, \sigma^2, \delta, \X^*_\ell ) =
\frac{f_{\vg}( \by^* |  \betab, \sigma^2, \X^*_\ell)^{1/\delta}}{\int f_{\vg}( \by^* |  \betab, \sigma^2, \X^*_\ell )^{1/\delta} \mathrm{d}\by^*}.
\label{density_normalized_power_lik} 
\end{equation} 
The above density-normalized power-likelihood is still a normal distribution
with variance inflated by a factor of $\delta$; in the above $\X_\vg^*$ denotes the imaginary design matrix under model $M_\ell$. 
In a similar manner, under the reference model, the density-normalized power-likelihood takes the form of \eqref{density_normalized_power_lik} but using now 
the likelihood $f_{0}( \by^* | \dn{\beta}_0, \sigma^2, \X^*_0) $ of $M_0$.


In order to apply the PEP methodology, the density-normalized power-likelihood (\ref{density_normalized_power_lik}) is used to evaluate, under the imaginary data and the baseline prior, the conditional prior predictive distribution 
$m_0^{\mathrm{N}}( \by^*|\sigma^2,\delta, \X^*_0)  $ of model $M_0$ 
as well as the conditional posterior distribution of $\dn{\beta}_\ell$
\begin{equation} 
\pi_{\vg}^{\N} ( \dn{\beta}_\ell | \by^*, \sigma^2, \delta, \X^*_\ell) 
= \frac{f_{\vg}( \by^* | \dn{\beta}_\ell, \sigma^2, \delta, \X^*_\ell)  \pi_{\ell}^{\mathrm{N}}(\dn{\beta}_\ell\,|\sigma^2, \X^*_\ell)}{m_\ell^{\mathrm{N}}( \by^*|\sigma^2,\delta, \X^*_\ell)},
\label{poste} 
\end{equation} 
where 
\begin{eqnarray} 
m_j^{\mathrm{N}}( \by^*|\sigma^2,\delta, \X^*_j) 
&=&\int f_j( \by^* | \dn{\beta}_j, \sigma^2, \delta , \X^*_j) \pi_j^{\mathrm{N}}(\dn{\beta}_j\,|\sigma^2, \X^*_j) \mathrm{d}\dn{\beta}_j, 
\label{priorell} 
\end{eqnarray} 
is the conditional prior predictive distribution of model $M_j$ for $j=\ell, 0$.

Finally, the  imposed prior for the parameters of any model $M_\ell$ has the following hierarchical structure 
\begin{equation}
\pi^{\mathrm{PEP}}_{\vg} ( \dn{\beta}_\ell, \sigma^2| \delta , \X^*_\ell) 
= \pi^{\mathrm{PEP}}_{\vg} ( \dn{\beta}_\ell | \sigma^2, \delta, \X^*_\ell) \pi^{\mathrm{N}}( \sigma^2 | \X^*_\ell),
\label{PEP-all}
\end{equation} 
with 
\begin{eqnarray}
\pi^{\mathrm{PEP}}_{\vg} ( \dn{\beta}_\ell | \sigma^2, \delta, \X^*_\ell) 
&=&\int \pi_{\vg}^{\N} ( \dn{\beta}_\ell | \by^*, \sigma^2, \delta, \X^*_\ell) m_0^{\mathrm{N}}(\by^* |\sigma^2,\delta , \X^*_0) \mathrm{d}\by^*. 
\label{PEP} 
\end{eqnarray}

The default choice for $\delta$ is to set it equal to $n^*$, i.e. the sample size of the imaginary data, so that the overall information of the imaginary data in the posterior is equal to one data point.
Furthermore, setting
$n^*=n$ and, consequently, the design matrix of the imaginary data $\X_\vg^*\equiv\X_\vg$ simplifies significantly the overwhelming computations required when considering all possible ``minimal'' training samples \citep{perez_berger_2002} while it also avoids the complicated issue (in some cases) of defining the size of the minimal training samples \citep{berger_pericchi_2004}. 
In addition, under the choice $n^*=n$, the PEP prior remains relatively non-informative even  for models with dimension close to the sample size $n$, while the effect on the evaluation of each model is minimal since the resulting Bayes factors are robust over different values of $n^*$. Detailed information about the default specifications of the PEP prior is provided in \cite{fouskakis_etal_2015}. 
Finally, the null model (with no explanatory variables) is a standard choice for the reference model in regression problems; see, for example \cite{perez_berger_2002}. 
 
In the rest of the paper, we briefly introduce the two PEP variants, under the normal linear model case, and we study and compare their properties by 
examining their dispersion and sparsity as well as the consistency of the induced model selection procedures.

\subsection{PEP prior variants} \label{variants}

\cite{Fouskakis_etal_2018_glm} introduced two alternative definitions of the PEP prior under the generalized linear model case. 
The core idea is to use the unnormalized power-likelihood $f_{\vg}( \by^* | \betab, \sigma^2, \X^*_\ell)^{1/\delta}$ and normalize the posterior density instead. 
This was also the approach followed by \cite{ibrahim_chen_2000} and \citet[Eq.4]{Friel_Pettitt_2008}. 
Under this view, the posterior distribution inside the integral of \eqref{PEP} is now derived as
\begin{eqnarray}
\pi_{\vg}^{\N, {\mathcal U}}( \betab | \by^*, \sigma^2,  \delta, \X^*_\ell )  
= \frac{     f_{\vg}( \by^*|\betab, \sigma^2, \X^*_\ell )^{\invd}\pi_{\vg}^{\N}(\betab | \sigma^2, \X^*_\ell)                 }
       {\int f_{\vg}( \by^*|\betab, \sigma^2, \X^*_\ell  )^{\invd}\pi_{\vg}^{\N}(\betab | \sigma^2, \X^*_\ell)\mathrm{d}\betab }.
\label{power_posterior}
\end{eqnarray} 

As a subsequent of using \eqref{power_posterior} in \eqref{PEP}, 
 two alternative PEP variants were introduced in \cite{Fouskakis_etal_2018_glm}:  a concentrated and a (more) diffuse version of PEP.  
The difference lies in the second component of the PEP definition given by \eqref{PEP}, 
that is, the predictive distribution $m_0^{\N}(\by^*|\sigma^2, \delta, \X^*_0)$ of $M_0$ which is used to average the posterior distribution \eqref{power_posterior}. 

In the concentrated-reference PEP (CR-PEP) we consider the usual prior predictive distribution 
$m_0^{\N}(\by^*|\sigma^2, \X^*_0)$ of $M_0$ given by \eqref{priorell} for $\ell=0$ and $\delta=1$, that is,  
\be
  m_0^{\N, {\mathcal CR}}(\by^*|\sigma^2, \delta, \X^*_0) =
  \int f_0(\by^*|\beta_0, \sigma^2, \X^*_0 )   \pi_0^{\N}(\beta_0 | \sigma^2, \X^*_0) \mathrm{d}\beta_0 .
\label{predictiveCR}
\ee 
 This approach adjusts the posterior distribution to account for $n^*/\delta$ data points but 
this (adjusted/power) posterior is averaged by using data from the actual predictive distribution of $M_0$ 
using data of size $n^*$.  

On the other hand, for the construction of the diffuse-reference PEP (DR-PEP), we consider the prior predictive distribution of $M_0$ based on the unnormalized power-likelihood. 
This makes the approach more diffuse than CR-PEP in the sense 
that also the predictive distribution results from a sample of size $n^*/\delta$. 
Hence, in DR-PEP the predictive distribution used in \eqref{PEP} is given by 
\be
  m_0^{\N, {\mathcal DR}}(\by^*|\sigma^2, \delta, \X^*_0) \propto 
  \int f_0(\by^*|\beta_0, \sigma^2, \X^*_0 )^{\invd}  \pi_0^{\N}(\beta_0 | \sigma^2, \X^*_0) \mathrm{d}\beta_0
\label{predictiveDR}
\ee 
where the above quantity is normalized in order $m_0^{\N, {\mathcal DR}}(\by^*|\sigma^2, \delta, \X^*_0) $ to be 
a probability density function in terms of $\by^*$.

Hence the definition of the two versions of PEP priors can be summarized as: 
\begin{eqnarray}
\pi_{\vg}^{\mathrm{VR-PEP}}(\betab| \sigma^2, \delta, \X^*_\ell)
&=& \int \pi_{\vg}^{\N, {\mathcal U}}( \betab | \by^*,  \sigma^2, \delta, \X^*_\ell )  
m_0^{\N, {\mathcal VR}}(\by^*|\sigma^2, \delta, \X^*_0) \mathrm{d}\by^* 
\label{VRPEP} 
\end{eqnarray}
with $\mathrm{VR} \in \{\mathrm{CR}, \mathrm{DR}\}$, 
$\pi_{\vg}^{\N, {\mathcal U}}( \betab | \by^*,  \sigma^2, \delta, \X^*_\ell )$ 
 defined in \eqref{power_posterior},  
$m_0^{\N, {\mathcal DR}}(\by^*|\sigma^2, \delta, \X^*_0)$ given by \eqref{predictiveDR} 
and $m_0^{\N, {\mathcal CR}}(\by^*|\sigma^2, \delta, \X^*_0) 
= m_0^{\N }(\by^*|\sigma^2,  \X^*_0)$ (see Eq. \ref{predictiveCR}). 

\begin{figure}[h!]
	\centering{}
	\includegraphics[scale=0.65]{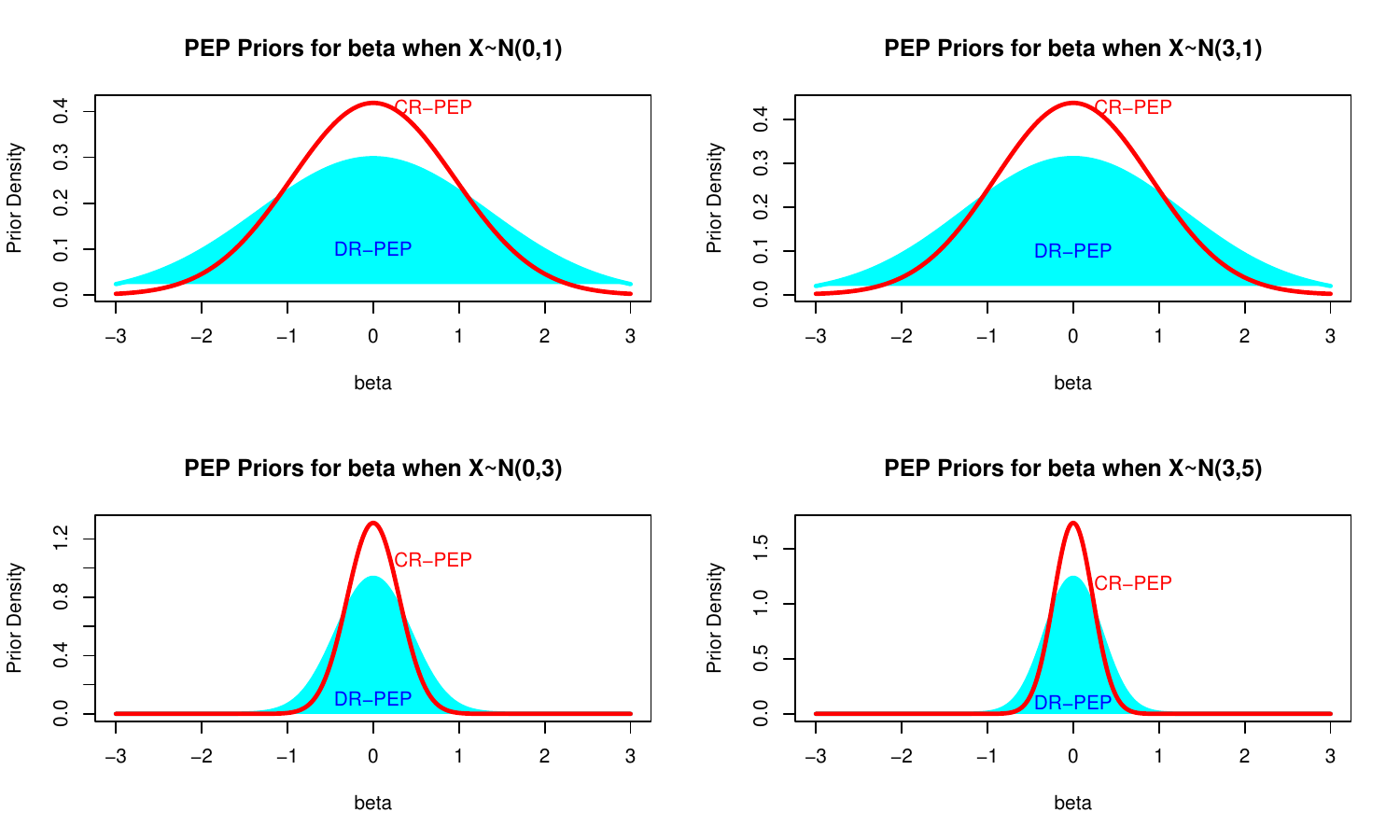}
	\caption{The marginal DR-PEP and CR-PEP priors, conditional on $\sigma=1$, using a simple normal linear regression model.}
	\label{plot1}
\end{figure} 

The above priors will be well defined under similar assumptions as in  \cite{perez_berger_2002}. 
Furthermore, impropriety of the baseline priors does not cause indeterminacy to the resulting Bayes factors, 
since $\pi_{\vg}^{\CR}(\betab| \sigma^2 ,\delta, \X^*_\ell)$ depends only on the normalizing constant of the baseline prior of the parameter of the null model 
and in $\pi_{\vg}^{\DR}(\betab| \sigma^2, \delta, \X^*_\ell)$ the normalizing constants cancel out.

In order to visualize the two PEP priors in (\ref{VRPEP}), we consider a simple normal linear regression case. Plots are given for the two marginal versions of PEP priors on the coefficient of the covariate $\mt{X}=\mt{X^*}$. We use $\sigma=1$, $\delta=n^*=100$ and the values of the explanatory variable are drawn from four different normal distributions. From Figure \ref{plot1} we see, in all cases, that both priors are centred at zero and the DR-PEP is more dispersed. Both priors are less dispersed as the variance of $\mt{X}$ increases.

\section{Properties of PEP variants  in normal regression}
\label{new_pep_normal_comparisons}

In this section we examine the properties of the DR-PEP and CR-PEP priors and compare them to the corresponding properties of the original PEP prior.
We work within the conjugate framework considered in \cite{Fouskakis_Ntzoufras_2016_jcgs}; specifically, we use as baseline priors a Zellner's $g$-prior for $\betab$ conditional on $\sigma^2$ and a reference prior for $\sigma^2$, that is  
$$
\pi_{\vg}^{\N}( \betab | \sigma^2, \X^*_\ell)=f_{N_{d_{\vg}}} \big( \betab ;  \mathbf{0}, g_0  ( \X_{\vg}^{*T} \X_{\vg}^* )^{-1} \sigma^2 \big)
\mbox{~and~ } \pi^{\N}(\sigma^2)\propto\sigma^{-2},
$$
where $d_{\ell}$ is the dimension of $\betab$ and 
$f_{N_k} \big( \cdot\,; \, \dn{\mu}, \, \dn{\Sigma} \big)$ is the density function of the $k$--dimensional multivariate normal distribution with mean vector $ \dn{\mu}$ and variance-covariance matrix $\dn{\Sigma}$.
In the following we use the default values for the hyperparameters discussed in Section \ref{sec1_intro}, namely $\delta=n$, $\nstar=n$, $\X_{\vg}^{*}=\X_{\vg}$. In addition, following \cite{Fouskakis_Ntzoufras_2016_jcgs} we set $g_0=n^2$; 
this way, the overall contribution of the PEP prior to the posterior will be equal to $(1+1/n)$ data points, corresponding to 
one point contributed from the power-likelihood part and $1/n$ from the baseline $g$-prior. 
As a reference model $M_0$ we consider the simplest nested model under consideration.

\subsection{Power-posterior component in PEP variants} 

Under both approaches and for any given model $M_{\ell}$, it is straightforward to show that the unnormalized likelihood is given by
\begin{eqnarray}
f_{\vg} \big(\by^* \, \big| \, \betab,\sigma^{2}, \X_\ell \big)^{\invd} & = & 
f_{N_{n}}\big(\by^*; \, \XG\betab, \, \sigma^{2}\I \big)^{\invd}\nonumber \\ 
& = &
\delta^{\frac{n}{2}}
(2\pi\sigma^{2})^{\frac{n(\delta-1)}{2\delta}}
f_{N_{n}}(\by^*;\XG\betab,\sigma^{2}\delta\I).
\label{a21}
\end{eqnarray}

Therefore, for both DR-PEP and CR-PEP priors, the posterior distribution \eqref{power_posterior}, conditional on the imaginary data, is given by 
\begin{eqnarray}
\pi_{\vg}^{\N, {\mathcal U}} \big( \betab \, | \, \by^*, \sigma^2, \delta, \X_\ell \big) & \propto &
f_{\vg} \big( \by^* \, \big| \, \betab, \sigma^2, \X_\ell \big)^{\invd}\pi_{\vg}^{\N}\big(\betab \, | \, \sigma^2, \X_\ell\big), \nonumber \\
& \propto & 
f_{N_{n}}\big(\by^*; \, \XG\betab, \,  \I \delta\sigma^{2} \, \big)
f_{N_{d_{{\vg}}}}\Big(\betab; \, \mathbf0, \, g_0(\XT\XG)^{-1} \sigma^{2}  \Big) \nonumber \\ 
& = & 
f_{N_{d_{{\vg}}}} \big(\betab; \, w \betahatstar, \, w\delta(\XT\XG)^{-1}\sigma^{2} \big),
\label{baseline_posterior}
\end{eqnarray} 
where $w=g_0/(g_0+\delta)$ and $\betahatstar$ is the maximum likelihood estimate based on the imaginary response $\by^*$. 
Thus, the posterior distribution 
involved in \eqref{VRPEP} is identical to the corresponding posterior under the original conditional PEP prior; see Equation 3 in \cite{Fouskakis_Ntzoufras_2016_jcgs}.

\subsection{Prior distributions and dispersion} 
\label{dispersion}

In this section we examine the volume of the variance covariance matrix of the two new versions of PEP priors and we compare them with the one under the $g$-prior.
This is important due to the connection of the volume of the variance with the dimensionality penalty induced in the Bayes factor for each pairwise model comparison; see for example \cite{dellaportas_etal_2012}. In short, the largest this volume is the highest the imposed penalty gets resulting, in its extreme form, to Lindley's paradox.   

\subsubsection{Diffuse-reference PEP prior} 
\label{drpepp}

For the DR-PEP setup, the prior predictive distribution of the imaginary data under model $M_{\ell}$ is given by 
$$
m_{{\vg}}^{\N, {\mathcal DR}}(\by^*|\sigma^{2},\delta, \X_\ell)  = 
\frac{m_{{\vg}}^{\U, {\mathcal DR}}(\by^*|\sigma^{2},\delta, \X_\ell) }
{\int m_{{\vg}}^{\U, {\mathcal DR}}(\by^*|\sigma^{2},\delta, \X_\ell) d\by^* },
$$
where $m_{{\vg}}^{\U, {\mathcal DR}}(\by^*|\sigma^{2},\delta, \X_\ell)$ is the normalizing constant of the power-posterior in \eqref{power_posterior} which is derived as follows
\begin{eqnarray}
m_{{\vg}}^{\U}(\by^*|\sigma^{2},\delta, \X_\ell) & = & 
\int f_{\vg}(\by^*|\betab,\sigma^{2}, \X_\ell)^{\invd} 
\pi_{{\vg}}^{\N}(\betab|\sigma^{2}, \X_\ell)
\mathrm{d}\betab\nonumber 
\\ 
& = &  \delta^{\frac{n}{2}}
(2\pi\sigma^{2})^{\frac{n(\delta-1)}{2\delta}} \times \nonumber\\
&  & \times 
\int
f_{N_{n}}(\by^*;\XG\betab,\sigma^{2}\delta\I)
f_{N_{d_{{\vg}}}}(\betab;\mathbf0,g_0(\XT\XG)^{-1}\sigma^{2})
\mathrm{d}\betab \nonumber
\\ 
& = & 
\delta^{\frac{n}{2}}
(2\pi\sigma^{2})^{\frac{n(\delta-1)}{2\delta}}
f_{N_{n}}(\by^*;\mathbf 0,\invLamda \sigma^{2}),
\label{a22}
\end{eqnarray}
with
\begin{equation*}
\invLamda = \delta \I+g_0\XG(\XT\XG)^{-1}\XT
\mbox{ and } 
\Lamda = \delta^{-1} \Big(\I-w\XG(\XT\XG)^{-1}\XT\Big). 
\end{equation*}
From the previous equations, it immediately follows that 
\begin{eqnarray}
m_0^{\N, {\mathcal DR}}(\by^*|\sigma^{2},\delta, \X_0) 
& = & f_{\mathrm{N}_{n}} \big(\by^*; \, \mathbf 0, \, \invLamdaO \sigma^{2} \big)
\label{predictiveDRnormal}
\end{eqnarray}
with
\begin{equation}
\invLamdaO
= \delta \I+ \tfrac{g_0}{n}\IotaN\IotaN^T
\mbox{ and } 
\LamdaO
= \delta^{-1} \left(\I-\tfrac{w}{n}\IotaN\IotaN^T \right). 
\label{lambda0}
\end{equation}

Thus, both components of the DR-PEP prior, that is the power-posterior in \eqref{baseline_posterior} and the prior predictive in \eqref{predictiveDRnormal}, are exactly the same as the corresponding components of the conditional PEP approach of \cite{Fouskakis_Ntzoufras_2016_jcgs}, where the density-normalized likelihood in \eqref{density_normalized_power_lik} was used. Hence, for Gaussian linear models the DR-PEP prior coincides to the original version of the conditional PEP and is given by 
\begin{eqnarray}
\pi_{\vg}^{\DR} \big(\betab| \, \sigma^2 \,\delta, \X_\ell \big) 
& =& f_{N_{d_{\vg}}}\big(\betab; \, \mathbf{0}, \, \V\sigma^2 \big) , 
     \label{PEPnormal} \\ 
\V &=& \delta\left( \XT\left[ w^{-1}\I-(\delta\LamdaO+w\HL)^{-1} \right]\XG\right)^{-1}		\nonumber  
\end{eqnarray}
with $\HL = \XG(\XT\XG)^{-1}\XT \nonumber$ and $\LamdaO$ given in \eqref{lambda0}.

The volume of dispersion of the DR-PEP prior is given by the determinant of the covariance matrix $\V$  
and equals 
\begin{equation}
\big | \V \big | =  \xi \times \big | \XT\XG \big |^{-1} 
\mbox{ with } 
\xi = \big\{ \delta w (w+1) \big\} ^{d_{\vg}-d_0}g^{d_0}.
\label{determinant}
\end{equation}
For the default values $\delta=n$ and $g_0=n^2$,  the variance multiplier $\xi$ appearing in \eqref{determinant} is equal to
\begin{equation}
\xi=n^{2d_{\vg}} \left[ \frac{2n+1}{(n+1)^2} \right] ^{d_{\vg}-d_0} > n^{d_{\vg}},  
\label{PEP_multiplier}
\end{equation}
where on the right hand side of the inequality we have the corresponding variance multiplier of Zellner's unit-information $g$-prior. The result in \eqref{PEP_multiplier} holds since 
\begin{eqnarray} 
\phi(n) &=& \log \xi - d_{\vg} \log n \nonumber \\ 
        &=& d_{\vg} \log n + (d_{\vg}-d_0) \log \left[\frac{2n+1}{(n+1)^2}\right]
\label{DRratio}
\end{eqnarray} 
is an increasing function of $n$ and $\phi(2)>0$; see \cite{Fouskakis_Ntzoufras_2016_jcgs} for details. 
Hence, the DR-PEP prior is more dispersed than Zellner's $g$-prior with $g=n$, for any sample size $n\ge2$, and consequently it leads to a more parsimonious variable selection procedure.

\subsubsection{Concentrated-reference PEP prior}
\label{sec_dispersion_CRPEP}

Under the CR-PEP approach, the prior predictive of the imaginary data, under the reference model, is given by 
\begin{eqnarray}
m_0^{\N, {\mathcal CR}}(\by^*|\sigma^{2}, \X_0) 
&  =  &  
f_{\mathrm{N}_{n}}(\by^*;\mathbf 0,\invLamdaOCR \sigma^{2}),
\label{prior_predictiveCR} 
\end{eqnarray}
with  
\begin{equation*}
\invLamdaOCR 
=  \I+g_0{n}^{-1}\IotaN\IotaN^T
\mbox{ and } 
\LamdaOCR 
=\I-\frac{g_0}{g_0+1}{n}^{-1}\IotaN\IotaN^T.
\end{equation*}
Combining \eqref{baseline_posterior} and \eqref{prior_predictiveCR}, we obtain the CR-PEP prior which has the same form as the DR-PEP in \eqref{PEPnormal} but with variance-covariance matrix 
\begin{equation*}
\VCR=\delta\left(
\XT\left[
w^{-1}\I-(\delta\LamdaOCR+w\HL)^{-1}
\right]\XG\right)^{-1}. 
\end{equation*}

As seen, the CR-PEP and DR-PEP priors differ only with respect to the variance-covariance matrix $\V$ appearing in \eqref{PEPnormal} where $\LamdaO$ is substituted by $\LamdaOCR$.
The volume of dispersion is now given by 
\begin{equation}
|\VCR| =  \xi \times |\XT\XG|^{-1} \mbox{ with } \xi= w^{d_{\vg}}\left({w+\delta}\right)^{d_{\vg}-d_0}\left(w+\delta+ wg_0\right)^{d_0}. 
\label{CRvolume} 
\end{equation}
For the derivation of the result in \eqref{CRvolume} see  \ref{appendix_CRvolume}.   
Under the default setting $\delta=n$ and $g_0=n^2$,  the variance multiplier becomes
\begin{align}
\xi & 
      = n^{2d_{\vg}}\left[\frac{n+2}{(n+1)^2}\right]^{d_{\vg}} \left[\frac{n^2+n+2}{n+2}\right]^{d_0}.
\label{CRmultiplier}
\end{align}
The log-ratio of the variance multipliers of the CR-PEP prior and the unit-information $g$-prior is given by 
\begin{eqnarray}
\phi(n) &=& \log \xi - d_{\vg} \log n \nonumber \\ 
&=& d_{\vg}\log\left(\frac{n^2+2n}{n^2+2n+1}\right)+d_0\log\left(\frac{n^2+n+2}{n+2}\right).
\label{CRratio}
\end{eqnarray}
For any model $M_{\ell} \supset M_0$ and under the restriction $1\le d_0<d_{\ell}\le n-1$
we obtain
\begin{eqnarray}
d_{\vg}\log\left(\frac{n^2+2n}{n^2+2n+1}\right)+d_0\log\left(\frac{n^2+n+2}{n+2}\right) &\ge&   \nonumber \\ 
(n-1)\log\left(\frac{n^2+2n}{n^2+2n+1}\right)+ \log\left(\frac{n^2+n+2}{n+2}\right)&=&\phi^*(n),
\label{CRminimum}
\end{eqnarray}
for $d_0 \in [1,n-2]$, $d_{\ell}\in[d_0+1,n-1]$ and any $n\in \{d_\ell+1, d_\ell +2, \ldots \}$. It can be proved that $\phi^*(n)$ is an increasing function of $n$ and additionally $\phi^*(2) > 0$ and thus $\phi^*(n)$ is always positive. Therefore, the log-ratio of the variance multipliers in \eqref{CRratio} will also be positive.   
Thus, the variance of the CR-PEP prior is larger than that of the $g$-prior, which means that CR-PEP prior will in general tend to favour less complex models. 
Additionally,  by rewriting the variance multiplier in \eqref{CRmultiplier} as  
\begin{eqnarray}
 \xi & = & n^{d_{\vg}}\left[\frac{n^2+2n}{n^2+2n+1}\right]^{d_{\vg}}
\left[\frac{n^2+n+2}{n+2}\right]^{d_0},
\label{CRmultiplier2}
\end{eqnarray}
we can see that for relatively large $n$ the first fraction in \eqref{CRmultiplier2} tends to one while the second fraction tends to $n$. Assuming that $d_0=1$, the CR-PEP variance multiplier is then approximately equal to $n^{d_{\vg}+1}$ and the log-ratio in  \eqref{CRratio} will be $\phi(n)\approx\log(n^{d_{\vg}+1}/n^{d_{\vg}})=\log(n)$. When the reference model $M_0$ is not the null model, the corresponding approximation is equal to $d_0\log(n)$.

The comparison with respect to the DR-PEP prior, and consequently to the original conditional PEP approach, is more straightforward. In this case, the log-ratio of the CR-PEP variance multiplier \eqref{CRmultiplier2} over the corresponding multiplier of the DR-PEP prior, given in \eqref{PEP_multiplier}, is
\begin{eqnarray}
\varphi(n)
& = &
\log\left(
\left[\frac{n+2}{(n+1)^2}\right]^{d_{\vg}}
\left[\frac{n^2+n+2}{n+2}\right]^{d_0}
\left[\frac{2n+1}{(n+1)^2}\right]^{d_0-d_{\vg}} 
\right) \nonumber \\
& = & 
\log\left(
\left[n+2\right]^{d_{\vg}-d_0}
\left[2n+1\right]^{d_0-d_{\vg}} 
\left[\frac{n^2+n+2}{(n+1)^2}\right]^{d_0}
\right)\nonumber \\
& = &
\log\left(
\left[\frac{n+2}{2n+1}\right]^{d_{\vg}-d_0}
\left[\frac{n^2+n+2}{n^2+2n+1}\right]^{d_0}
\right).
\label{ratioPEP}
\end{eqnarray}
Both fractions appearing in \eqref{ratioPEP} are equal to or smaller than one for any $n\ge 1$.
Therefore, the log-ratio is always negative. 
This implies that the CR-PEP prior induces a variable selection procedure which is less parsimonious than the corresponding one under the DR-PEP prior.

\begin{figure}[h!]
	\centering{}
	\psfrag{dl = 5}[c][c][1.0]{ $d_\ell=5$ }
	\psfrag{dl = 10}[c][c][1.0]{ $d_\ell=10$ }
	\psfrag{dl = 50}[c][c][1.0]{ $d_\ell=50$ }
	\psfrag{dl = 100}[c][c][1.0]{ $d_\ell=100$ }
	\psfragscanon 
	\psfrag{dl}[c][c][1.0]{ $d_\ell$ }
	\includegraphics[scale=0.77]{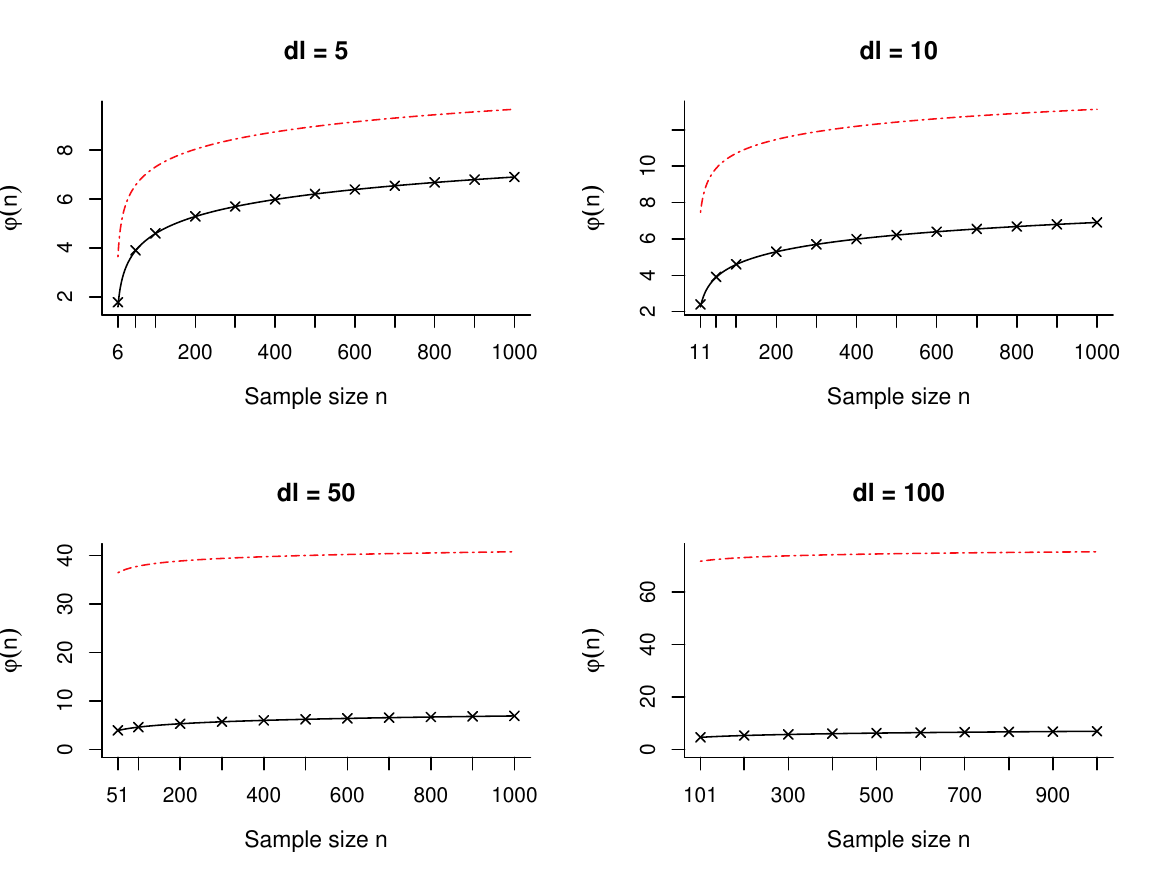}
	\caption{Log-ratios of the variance multipliers  $\varphi(n)$ of the DR-PEP prior (dashed red line) and CR-PEP prior (solid black line) over the unit-information $g$-prior for $d_{\vg}=5,10,50,100$ and varying sample size; the crosses correspond to the approximation $\log(n)$.}
	\label{Ratios}
\end{figure}

\subsubsection{Numerical illustrations}
\label{numerical}

Here we provide some basic illustrations that highlight the behaviour of the variance multipliers of the CR-PEP and DR-PEP priors for varying sample size and number of predictors, given the restriction that $n\ge d_{\vg}+1$ and assuming that $d_0=1$.

%
%

The log-ratios of the DR-PEP and CR-PEP prior multipliers over the unit-information $g$-prior multiplier  
(see respective Eqs. \ref{DRratio} and \ref{CRratio}), for increasing sample size $n$ and selected values of $d_{\vg} \in \{ 5, 10, 50, 100 \}$, are illustrated in Figure \ref{Ratios}.
For both prior setups, the log-ratios are positive and increasing with the sample size $n$, with  
the DR-PEP being always more dispersed than the CR-PEP as expected according to Section \ref{sec_dispersion_CRPEP}.  
Additionally, the log-ratio of the DR-PEP prior over the $g$-prior increases as $d_{\vg}$ gets larger, whereas the ratio of the CR-PEP prior over the $g$-prior is not affected by $d_{\vg}$ as it remains constant, approximately equal to $\log(n)$. 
 
\begin{figure}[h!]
	\centering{}
	\psfrag{dl = 10}[c][c][1.0]{ $d_\ell=10$ }
	\psfrag{dl = 25}[c][c][1.0]{ $d_\ell=25$ }
	\psfrag{dl = 50}[c][c][1.0]{ $d_\ell=50$ }
	\psfrag{dl = 100}[c][c][1.0]{ $d_\ell=100$ }
	\includegraphics[scale=0.77]{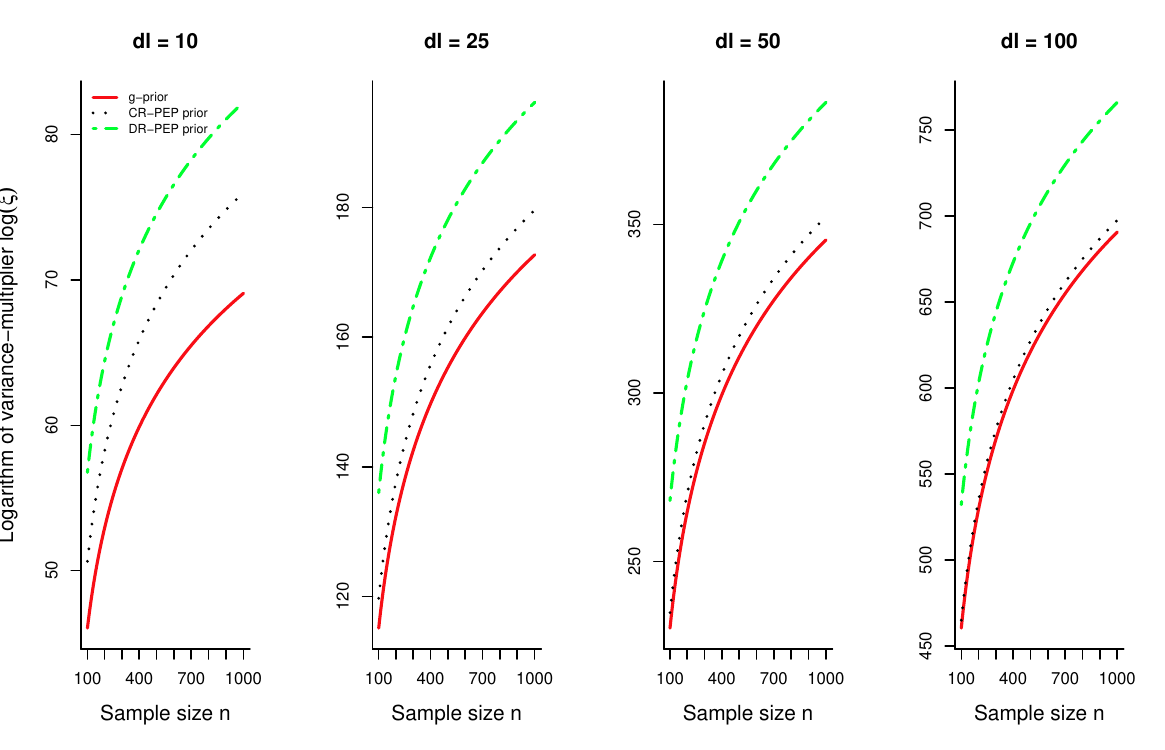} \\
	\vspace{-1em}
	\caption{Log-scaled plots of the variance multipliers ($\xi$) of the DR-PEP, CR-PEP and Zellner's $g$ priors for $d_{\vg}=10,25,50,100$ and sample size  from 100 to 1000.}
	\label{Multipliers1_plot}
\end{figure} 

In Figure \ref{Multipliers1_plot} we present on log-scale the variance multipliers of the CR-PEP, DR-PEP and the unit-information $g$ priors for $d_{\vg} \in \{10, 25, 50, 100 \}$ and sample size ranging from 101 (the minimum size required for $d_{\vg}=100$) to 1000.
As seen, as model dimensionality increases all priors become more dispersed; however, the distance between the variance multiplier of the DR-PEP prior and the corresponding multipliers of the CR-PEP and the $g$-prior is also increasing with $d_{\ell}$. Potentially, this feature of the DR-PEP prior makes it more suitable for problems involving a large number of predictors and where the aim is to have a parsimonious model selection method.

\subsection{Marginal likelihood and model selection consistency}
\label{cons}

In this Section we examine the limiting behaviour of the marginal likelihood of any model $M_\ell$, under the DR-PEP and the CR-PEP priors, under the M-closed scenario. We have assumed that $d_\ell$ 
does not increases with the sample size $n$ and additionally $d_\ell < n$.  

The posterior distribution of $\betab$ and $\sigma^2$ under either the conditional PEP prior of \cite{Fouskakis_Ntzoufras_2016_jcgs} or the DR-PEP prior, examined here,
is given by 
\begin{equation}
\pi_{\vg}^{\DRP}(\betab,\sigma^2|\by,\delta, \X_\ell) =
f_{N_{d_{\vg}}} \big (\betab ; \widetilde{\boldsymbol\beta}_{\vg} ,\SG\,\sigma^2 \big)
f_{IG} \big ({\sigma^2;\widetilde{a}_{\vg},\widetilde{b}_{\vg}} \big),
\label{PEPposterior}
\end{equation}
where $\widetilde{\boldsymbol\beta}_{\vg}=\SG\XT\by$, $\SG=\big (\V^{-1} +\XT\XG\big)^{-1}$, $\widetilde{a}_{\vg}=n/2$, $\widetilde{b}_{\vg}=\mathrm{SS}_{\vg}/2$ with
$\mathrm{SS}_{\vg}=\by^T\big ( \mathbf{I}_n+\XG\V\XT\big)^{-1}\by$, and $f_{IG}(\cdot\,;\,a,b)$ denotes the density of the inverse gamma distribution with shape parameter $a$ and scale parameter $b$. In the above, $\V$ is given in Section \ref{drpepp}.

Then, the marginal log-likelihood is given by
\begin{eqnarray}
\log m_{\vg}^{\DRP}(\by|\delta, \X_\ell) &=& C-\frac{1}{2}\log\big|\mathbf{I}_n+\XG\V\XT\big|-\nonumber\\ & & \frac{n}{2}\log \big( \by^T(\mathbf{I}_n+\XG\V\XT)^{-1}\by\big),
\label{logmargPEP}
\end{eqnarray}
where $C$ is a constant that does not depend on the structure of model $M_{\ell}$. 
For large $n$, the marginal log-likelihood in \eqref{logmargPEP} can be approximated by
\begin{equation}
\log m_{\vg}^{\DRP}(\by|\delta, \X_\ell) \approx C-\frac{1}{2}\mathrm{BIC}_{\vg}.
\label{BIC}
\end{equation}
Thus, the marginal likelihood under the DR-PEP prior has the same limiting behaviour as the BIC which is known to be consistent under a minor realistic assumption 
\citep{fernandez_etal_2001,liang_etal_2008}. For  a detailed proof of \eqref{BIC} see \cite{Fouskakis_Ntzoufras_2016_jcgs}.

\color{black}


Similarly to \eqref{logmargPEP}, the marginal log-likelihood under the CR-PEP prior is
\begin{eqnarray}
\log m_{\vg}^{\CR}(\by|\delta, \X_\ell) &=& C-\frac{1}{2}\log\big|\mathbf{I}_n+\XG\VCR\XT\big|-\nonumber \\
& & \frac{n}{2}\log \big( \by^T(\mathbf{I}_n+\XG\VCR\XT)^{-1}\by\big).
\label{margCR-PEP}  
\end{eqnarray}
Following the proof \citet[see Section D, Eqs. D.1--D.2 of the Supplementary Material]{Fouskakis_Ntzoufras_2016_jcgs}, 
the first logarithmic term  yields
\begin{eqnarray}
 |\I+\XG\VCR\XT| &=& (1+\delta w)^{d_\vg}|\LamdaOCR|^{-1}\left|\LamdaOCR+\left(\frac{w^2}{1+\delta w}\right)\HL\right| \nonumber \\ 
 &\approx&(1+\delta)^{d_\vg}|\LamdaOCR|^{-1}\left|\LamdaOCR+\left(\frac{1}{1+\delta}\right)\HL\right| \nonumber \\ 
  &\approx&(1+\delta)^{d_\vg}. 
\end{eqnarray}
Note that the approximation is accurate for large $n$ when $\delta=n$ and $g_0=n^2$, so that $w=g_0/(g_0+\delta)\approx1$. Given these values, we can also approximate the second logarithmic term in \eqref{margCR-PEP} by 
\begin{equation}
{\by}^T(\I+\XG\VCR\XT)^{-1}\by \approx {\by}^T\by-{\by}^T\XG\left(
\XT\XG\right)^{-1}\XT\by 
\equiv \mathrm{RSS}_{\vg}, 
\label{RSS_CR}
\end{equation}
where RSS$_{\vg}$ is the usual residual sum of squared of model $M_{\ell}$. The derivation for \eqref{RSS_CR} is provided in \ref{appendix_RSS_CR}.  
Hence, the marginal log-likelihood under the CR-PEP prior is approximately given by 
\begin{eqnarray}
\log m_{\vg}^{\CR}(\by|\delta, \X_\ell) & \approx & C-
\frac{d_{\vg}}{2}\log(n+1)
-\frac{n}{2}\log(\mathrm{RSS}_{\vg})\nonumber \\
& \approx & C-
\frac{d_{\vg}}{2}\log(n)
-\frac{n}{2}\log(\mathrm{RSS}_{\vg})\nonumber \\
&\approx &C-\frac{1}{2}\mathrm{BIC}_{\vg}, 
\end{eqnarray}
for $\delta=n$ and large $n$. Thus, variable selection, based on the CR-PEP prior with a $g$-prior as baseline, has also the same limiting behaviour as the BIC and is, therefore, consistent.  

Following a comment of a referee, we have calculated the standardized inner product between the likelihood and the prior which provides a geometric perspective of the marginal likelihood. 
This quantity was defined by \cite{carvalho_2019} and is given by 
$$
\kappa_{\pi_\ell,f_\ell} = \frac{ m_\ell(\by) }{ ||f_{\vg}( \by | \dn{ \beta }_\ell, \sigma^2,  \X_\ell ) ||~  
	|| \pi_\ell ( \dn{ \beta }_\ell, \sigma^2 ) ||  }, 
$$
where 
$|| g(\dn{\theta}) || = \smash{\sqrt{ \int g(\dn{\theta})^2 d\dn{\theta} }} $ 
for any function or distribution of the parameters under consideration $\dn{\theta}$ 
(here we consider the conflict/comparison between the prior and the likelihood). 
This measure provides a normalized version of the marginal likelihood and
measures the agreement between the prior and the data likelihood. 
When $\kappa_{\pi_\ell,f_\ell}$ equals to one, the prior is in total agreement with the likelihood, while when 
$\kappa_{\pi_\ell,f_\ell}\rightarrow 0$ there is a large conflict between the prior and the likelihood. 
In terms of model selection, the posterior Bayes factor of \cite{Aitkin_1991} essentially 
leads to $\kappa_{\pi_\ell,f_\ell}=1$ and generally it is not a desirable or realistic case in practice. 
On the other hand, within the objective Bayes framework, we wish to have a small $\kappa_{\pi_\ell,f_\ell}$ but well separated from zero since values very close to zero will indicate the activation of the 
\cites{bartlett_57} paradox. 
In our case we use an improper prior (Jeffrey’s prior) for the error variance $\sigma^2$, 
therefore, the above quantity
cannot be calculated due to the unknown normalizing constant.


\section{Simulation studies}
\label{sec:sim}
In this Section we present comparisons based on simulations in order to study the performance of the proposed methods under various scenarios. In Section \ref{S_model} we focus on one case of correct model specification and three cases of misspecified models, considering independent as well as correlated predictors and also increasing sample size. In Section \ref{S_p} we investigate the setting where the model is correctly specified but the number of covariates grows with sample size, assuming again that covariates are either independent or correlated. 

We compare DR-PEP and CR-PEP to ``well-established" priors routinely used for Bayesian variable selection; namely, the $g$-prior \citep{zellner_76}, the hyper-$g$ prior
\citep{liang_etal_2008} and the hyper-$g/n$ prior
\citep{liang_etal_2008}. All competing methods were implemented using  
the \texttt{BAS} package in \texttt{R}; we set $g = n$ in the $g$-prior in order to
implement  the unit-information prior \citep{kass_wasserman_95} and $\alpha = 3$ in
the hyper-$g$ and hyper-$g/n$ priors as recommended by
\cite{liang_etal_2008}. A beta-binomial prior on the model space, with both parameters equal to 1, is used. Each simulation described below is repeated 100 times. \texttt{R} reproducible code for all simulations is available at \url{https://github.com/kperrakis/PEP_variations}. 

\subsection{Simulations under M-open and M-closed scenarios}\label{S_model}
In these simulations we assume that dimensionality is $p=10$ with three out of the ten covariates ($X_1,\cdots,X_{10}$) actually relating to the response; specifically, the assumed true relationship between the response variable and the influential predictors is 
\begin{eqnarray} 
\label{ss}
Y_i = 0.3X_{i1} + 0.5X_{i3} + X_{i4} + \varepsilon_i,
\end{eqnarray}
for $i=1\cdots,n$ and $n \in \{30, 50, 100, 500, 1000\}$. 
Similarly to the setup in \cite{Rossell_Rubio2018} we consider the following four cases for the distribution of the random errors $\varepsilon_i$: (a) normal, (b) Laplace, (c) asymmetric normal, and (d) heteroscedastic normal. Case (a) is an M-closed case as the true model is included in the set of models under consideration; here the errors are generated as $\varepsilon_i\sim N(0,1)$. In contrary, cases (b) - (d) are all M-open scenarios. Specifically, in case (b) we have heavier tails than in case (a) with errors $\varepsilon_i\sim L(0,1/\sqrt{2})$, where $L(\mu,b)$ denotes the Laplace distribution with location $\mu$ and scale $b$; the latter parameter is set so that the variance equals one. In case (c) we utilize the asymmetric (or two-piece) normal distribution under the epsilon-skew parameterization \citep{Mudholkar_Hutson_2000}, denoted as $AN(\mu,\vartheta,\alpha)$, with location parameter $\mu\in\mathbb{R}$, scale parameter $\vartheta\in \mathbb{R}^+$ and asymmetry parameter $\alpha\in[-1,1]$. Its probability density function is 
\begin{equation}
f(x)=\frac{1}{\sqrt{2\pi\vartheta}}\Bigg[\exp\Bigg(-\frac{(x-\mu)^2}{\vartheta(1-\alpha)^2}\Bigg)I(x<\mu)+
\exp\Bigg(-\frac{(x-\mu)^2}{\vartheta(1+\alpha)^2}\Bigg)I(x\ge\mu)\Bigg],
\end{equation}
where $I(\cdot)$ is the indicator function and $x\in \mathbb{R}$. Here, we assume that $\varepsilon_i\sim AN(0,1,0.5)$ which gives approximately $\mathrm{Var}(\varepsilon_i)\approx 1.12$. Finally, for the heteroscedastic case (d) we simulate initially $\tilde{\varepsilon_i}\sim N(0,1)$ and then set $\varepsilon_i=\exp(0.3X_{i1} + 0.5X_{i3} + X_{i4})\tilde{\varepsilon_i}/c$, where $c$ is tuned in such a way so that $\mathrm{Var}(\varepsilon_i)=\mathrm{Var}(\tilde{\varepsilon}_i)$ in order to have comparable signal-to-noise ratio with the previous cases \citep{Rossell_Rubio2018}.        

\subsubsection{Scenario 1: Independent covariates} \label{S1}

\begin{figure}[h]
	\centering{}
	\vspace{-2em}
	\includegraphics[scale=0.6]{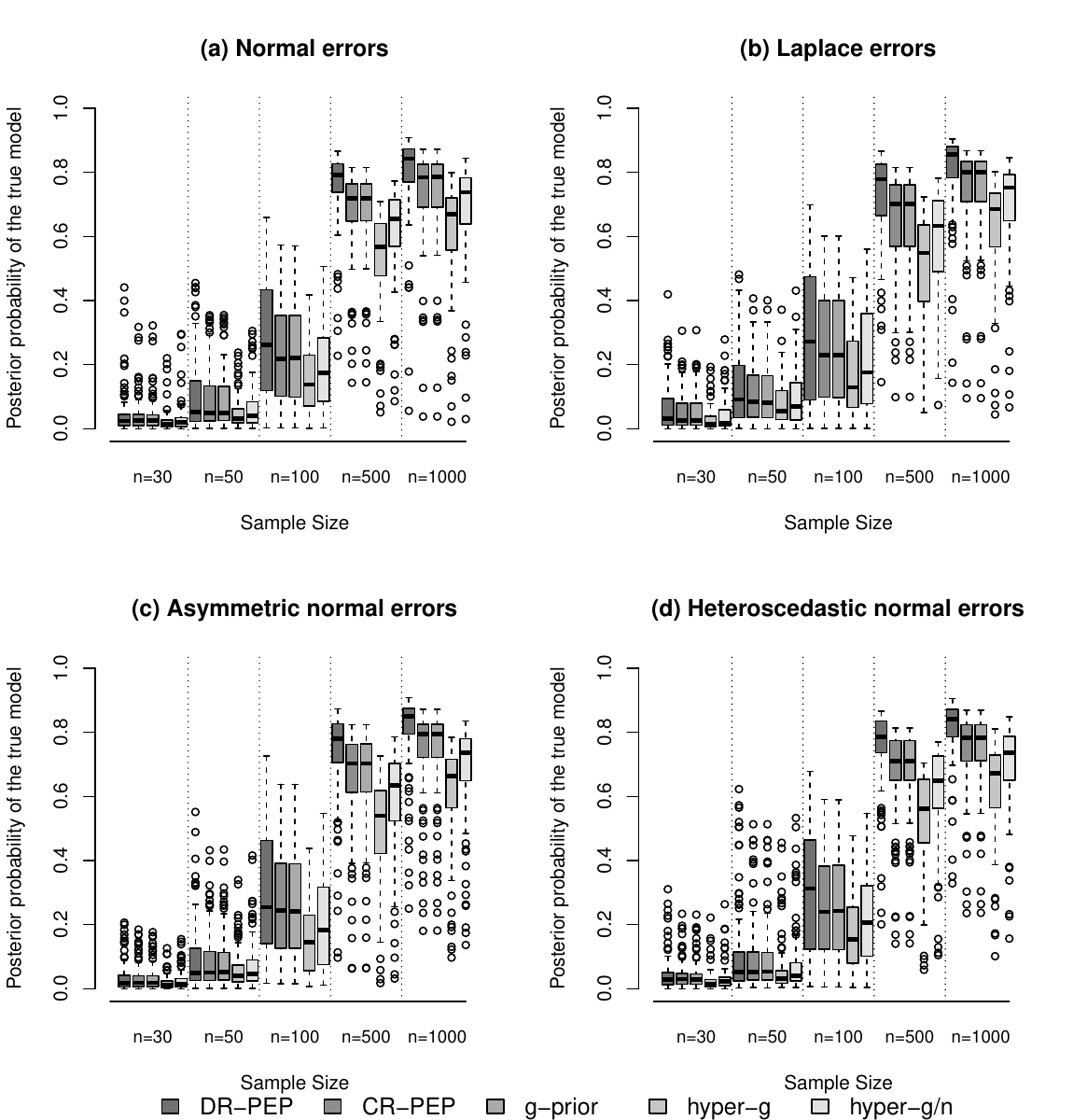}
	\vspace{-1em}
	\caption{Posterior  probabilities of the true model vs. sample size under the four model cases for Scenario 1 (independent covariates).}
	\label{plot_true_model}
\end{figure}

In this first scenario the covariates are generated independently from a Gaussian distribution.
Figure \ref{plot_true_model} depicts the between-samples distribution of the posterior probability of the true model for the Bayesian variable selection techniques 
under comparison and the four cases under consideration.
Focusing initially on the M-closed case (a), it is clear that for small sample sizes all methods under consideration fail to provide high posterior evidence in favor of the true model. 
As the sample size gets larger, all methods increase their posterior support towards the true model, 
with DR-PEP to perform slightly better than CR-PEP and the Zellner's $g$-prior, followed by the hyper-$g/n$ and the hyper-$g$ priors. 
This is sensible since the two proposed methods together with the Zellner's $g$-prior, are converging to the same Bayes factors as $n$ grows but 
with the proposed approaches constantly supporting more parsimonious models. 
On the other hand, the hyper-$g$ and the hyper-$g/n$ priors give the lowest support towards the true model 
due to their hierarchical structure which increases the posterior uncertainty on the model space. 
These two methods need larger sample size, than the rest of the approaches, in order to fully a-posteriori support the true generating mechanism. 
Interestingly, results are almost identical for the three M-open cases (b), (c) and (d); this provides some empirical evidence that the methods lead to robust model selection under the particular cases of model misspecification and when covariates are independent. 

\begin{figure}[h!]
	\centering{}
	\includegraphics[scale=0.67]{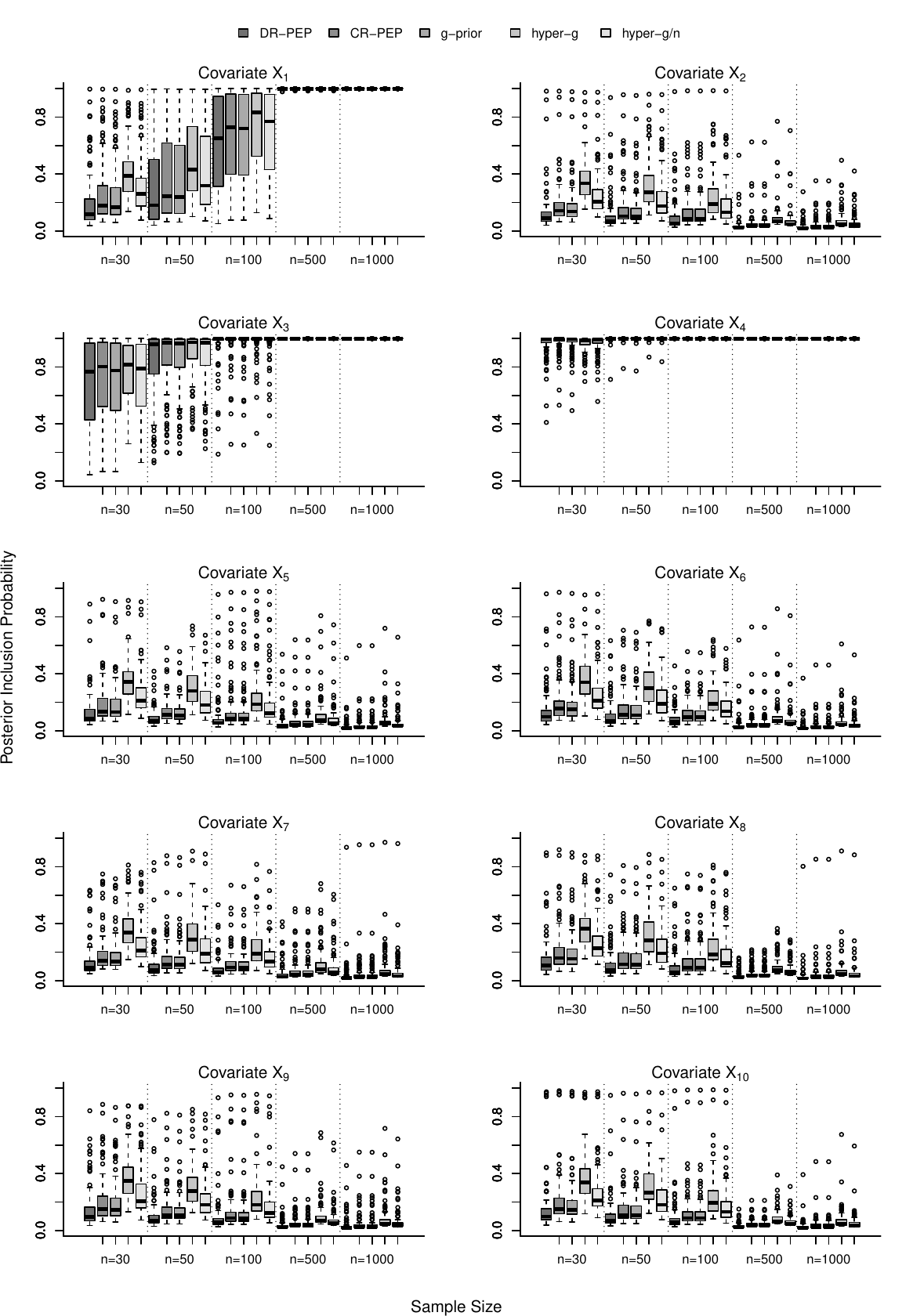}
	\caption{Posterior  inclusion probabilities of each covariate for Scenario 1 (independent covariates) under normal errors (M-closed case).}
	\label{inc_probs_ex1_normal}
\end{figure}

We now turn to inspection of posterior inclusion probabilities starting with case (a); these are presented in Figure \ref{inc_probs_ex1_normal}. 
First, we observe that all methods successfully identify $X_4$ (with true effect equal to one) as an important 
component of the model, even for small sample sizes. 
Furthermore, the between-samples variability of the posterior inclusion probabilities reduces as the sample size increases. 
Similar is the picture for the posterior inclusion probabilities of the other two covariates with non-zero effects, $X_1$ and $X_3$, 
but with slower rates of convergence towards one. 
For covariate $X_1$ (with true effect equal to $0.3$) we observe large between-samples 
uncertainty concerning the importance of this effect even with $n = 100$ under all methods. 
For $n\ge 500$, all methods successfully identify the importance of this covariate with almost zero between-samples variability. 
In general, the hyper-$g$ method supports this covariate with slightly higher inclusion probabilities, compared to the other methods; with DR-PEP we observe slightly lower inclusion probabilities, compared to the other approaches, for small $n$. 
This is  due to the characteristics of the two methods, with the first supporting more complicated models 
and the latter more parsimonious ones. 
Finally for covariate $X_3$ (with true effect equal to $0.5$), all methods successfully identify it as important with almost zero between-samples variability for $n\geq 100$. For smaller sample sizes the methods behave similarly, showing  large between-samples uncertainty concerning the importance of this effect and almost identical median values.

The between-samples distribution of the posterior inclusion probabilities for all covariates with zero true effects is similar; see Figure \ref{inc_probs_ex1_normal}. 
It is noticeable that all methods identify, fairly fast, that these covariates should have low posterior inclusion probabilities with the 
between-samples variability decreasing as $n$ gets larger. DR-PEP prior shows the best behaviour, followed by the CR-PEP and the $g$-prior that behave in a similar manner and then by the hyper-$g/n$ and hyper-$g$ priors.
In general the posterior inclusion probabilities under the  hyper-$g$ prior are on average higher under small sample sizes (close to $0.4$, for $n=30$ for example). 
This increases the posterior uncertainty on the model space and results 
to lower probabilities of identifying the true model as the maximum a-posteriori model. 
Furthermore, it is also noticeable that the inclusion probabilities under the hyper-$g$ and hyper-$g/n$ priors 
seem to converge slower towards zero as $n$ gets larger, both in terms of median values and in terms of between-samples variability.
To sum-up,  all methods identify the true model structure with increasing probability as $n$ increases. All methods tend to select simpler models than the true model structure when they fail while the covariates identified falsely as important are relatively low even for small samples for the PEP and the $g$ priors but considerably higher for the hyper-$g$ priors (especially for small samples).

The corresponding results for the M-open cases (b), (c) and (d) can be found in \ref{Sim1}; see Figures \ref{inc_probs_ex1_laplace}, \ref{inc_probs_ex1_a_normal} and \ref{inc_probs_ex1_h_normal}, respectively. 
Overall, we cannot detect significant differences with respect to the results obtained under case (a); a finding which is generally in line with Figure \ref{plot_true_model}.

To sum up, the main finding from this simulation is that the DR-PEP and CR-PEP methods identify the true model structure with (slightly) higher posterior probability than the rest of the methods, with the DR-PEP prior to behave slightly better. They both provide posterior inclusion probabilities close to zero for non-important effects (even for small sample sizes) and high inclusion probabilities for the important effects (although these are slightly smaller than the ones obtained under the hyper-$g$ and hyper-$g/n$ priors for small sample sizes).

\subsubsection{Scenario 2: Highly-correlated covariates} \label{S2}

In this second simulation the covariates are drawn from a multivariate normal and have again marginally a zero mean and a unit variance, but this time we introduce strong correlations amongst them; namely, all correlations are set equal to 0.9. It is well known that this causes problems in variable selection methods, especially under small samples; see for example \cite{Ghosh_Ghattas_2015}.

\begin{figure}[h]
	\centering{}
	\vspace{-1em}
	\includegraphics[scale=0.6]{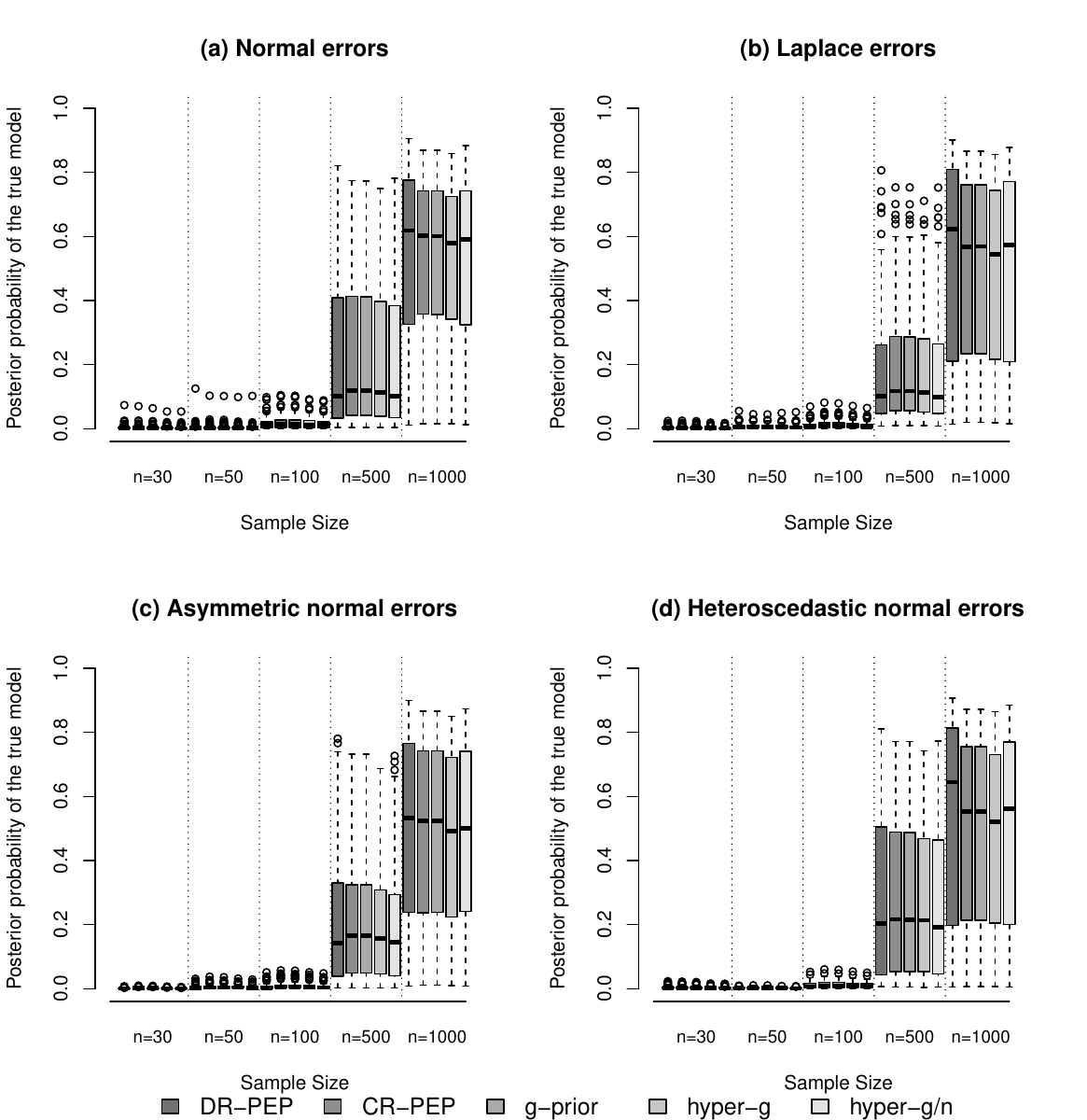}
	\caption{Posterior  probabilities of the true model vs. sample size under the four model cases for Scenario 2 (highly-correlated covariates).}
	\label{plot_true_model2}
\end{figure}

\begin{figure}[H]
	\centering{}
	\includegraphics[scale=0.67]{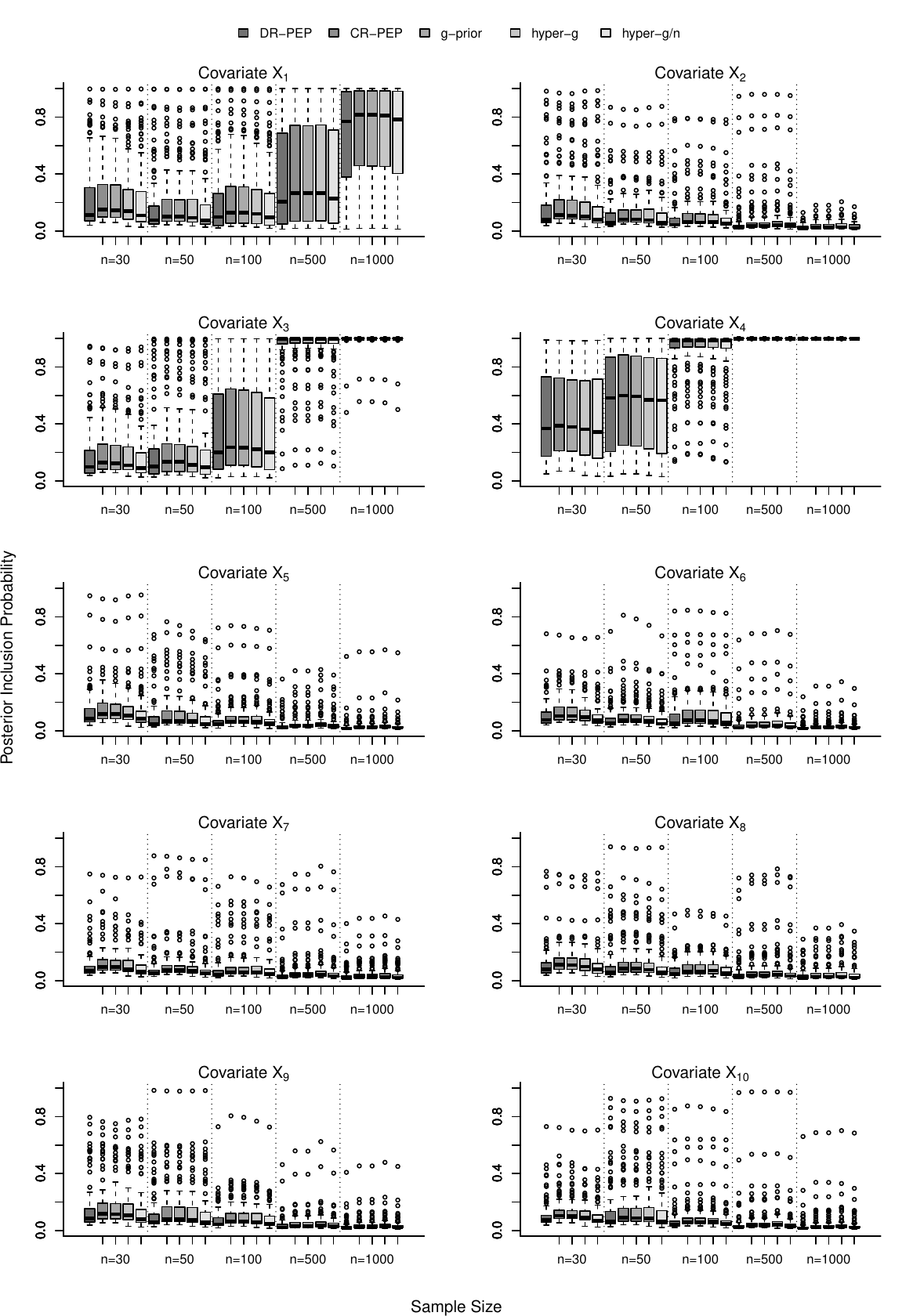}
	\caption{Posterior  inclusion probabilities of each covariate for Scenario 2 (highly-correlated covariates) under normal errors (M-closed case).}
	\label{inc_probs_ex2_normal}
\end{figure}

Again, we start our analysis with the resulting posterior probabilities of the true model, initially focusing on case (a) in Figure \ref{plot_true_model2}. 
In general, all methods perform similarly in this scenario; as expected a large sample size is needed for the identification of the true model due to the high correlations. Specifically, we see that all methods fail to provide high posterior evidence in favor of the true model even when $n=500$. For $n=1000$, posterior model probabilities are on average above 0.5 and all methods seem to perform equally good; however we still have high between-samples variability in the estimates.
The situation is again similar under the M-open cases (b), (c) and (d) in Figure \ref{plot_true_model2}; overall there are no significant different with reference to case (a). We observe only some slight differences; specifically, in cases (b) and (c) we have a small drop in the posterior probabilities under $n=500$ and, surprisingly, probabilities are a bit higher in case (d) under $n=500$ and $n=1000$.

The posterior inclusion probabilities under case (a) are presented in Figure \ref{inc_probs_ex2_normal}. All methods successfully identify the importance of covariate $X_4$ for $n=100$, with between-samples variability diminishing for larger $n$. In comparison, for covariate $X_3$ all methods require a larger sample size ($n=500$ in our experiments) in order to reach high posterior inclusion probabilities (close to one), while for $X_1$ (which has the smallest effect) convergence to one is much slower as we reach median inclusion probabilities above 0.5 only when $n=1000$, and even in this case between-samples variability is large.
Concerning the non-influential covariates which have a zero effect to the response,    
their estimated posterior inclusion probabilities are quite similar across competing methods. 
Specifically, all methods identify fairly fast, i.e. even under small sample sizes, that these covariates should have low posterior inclusion probabilities. 
To sum up,  all methods identify the true model with increasing posterior probability as $n$ grows but, in this scenario, larger sample size is required to identify the true model in comparison with the independent covariates case. 
Moreover,  here, zero coefficients are identified by all methods  even for small samples.  
Hence, the behavior of selecting more parsimonious models  is more intense in this case than in the scenario of Section \ref{S1}. 

The corresponding plots with the posterior inclusion probabilities under the three M-open cases can be found in  \ref{Sim2}; specifically, see Figure \ref{inc_probs_ex2_laplace} for case (b), Figure \ref{inc_probs_ex2_a_normal} for case (c) and Figure \ref{inc_probs_ex2_h_normal} for case (d). Again, results are more or less the same to those obtained in the M-closed case. Correlations seem to affect more the heteroscedastic case (d), where we observe higher variance than usual in certain estimates; see Figure \ref{inc_probs_ex2_h_normal}.

\subsection{Simulations with growing $p$ and $n$}
\label{S_p}

In this Section we focus on the M-closed case, allowing this time model dimension to increase with sample size.
In particular, we assume the same true linear effects as in Eq. \eqref{ss}, with all covariates either independent (as in Section \ref{S1}) or highly correlated (as in Section \ref{S2}), for $n\in\{30, 50, 100, 500, 750\}$  and $p =\lceil n^{0.4} \rceil \in\{ 4, 5, 7, 13, 15\}$; where $\lceil x \rceil$ denotes the least integer greater than or equal to the real number $x$. 
\begin{figure}[H]
	\centering{}
	\includegraphics[scale=0.75]{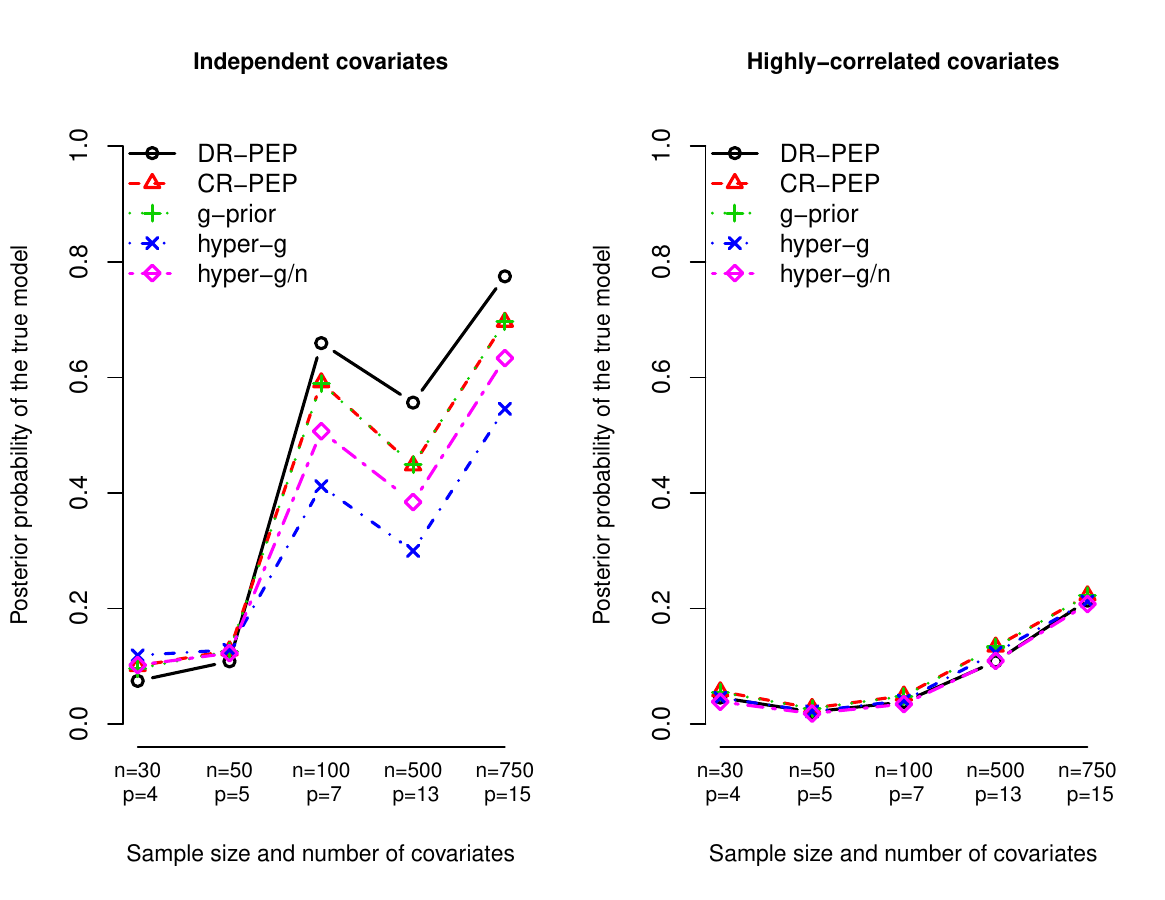}
	\caption{Posterior  probabilities of the true model vs. sample size and model dimensionality under independent covariates (left) and highly-correlated covariates (right).}
	\label{plot_grow_p_true_model}
\end{figure}

The posterior model probabilities of the true model are presented in Figure \ref{plot_grow_p_true_model}. Under independent covariates we see that with the DR-PEP prior we have a faster convergence to one, while the CR-PEP and $g$-prior (which exhibit identical behaviour) follow as second best. The differences in posterior model probabilities showcase the PEP properties, examined in this paper, in comparison to properties of the hyper-$g$ prior. Under highly-correlated predictors we have an entirely different picture; here, as expected, convergence to one is very slow, irrespective of the prior. All methods behave similarly, with the CR-PEP and $g$-prior having just a very slight edge. In general, Figure \ref{plot_grow_p_true_model} highlights the impact of high correlations on variable selection.

%

The resulting posterior inclusion probabilities (averages) from the independence and high-correlation scenarios are shown in Figures \ref{inc_probs3} and \ref{inc_probs4}, respectively, in \ref{AppSim3}. In the former case we generally observe similar patterns regarding the influential predictors, except for $X_1$ under ($n=30, p=4$) and ($n=50, p=5$); in this specific cases the hyper-$g$ has a higher detection rate for this predictor. We also notice that the DR-PEP is more effective in identifying the covariates with a zero effect.
In the second case (high-correlations), all methods are effective overall in detecting the importance of $X_4$, in the sense that this predictor would be included in the median probability model in all ($n,p$) settings under consideration. This is also the case for covariate $X_3$, but to a lesser extend; i.e. not when ($n=50,p=5$) and ($n=100,p=7$). Here the problematic case is that of predictor $X_1$ which has the smallest non-zero effect; as seen this covariate never enters the median probability model regardless of method. The failure in detecting $X_1$ also explains the very low posterior model probabilities (of the true model) witnessed in Figure \ref{plot_grow_p_true_model}.

%

 \section{Discussion}
\label{disc}  

In this paper we examined the properties of two new versions of the PEP prior, which have been recently proposed in the context of objective Bayesian variable selection \citep{Fouskakis_etal_2018_glm}, namely the CR-PEP and DR-PEP priors. 
Specifically, we compared the dispersion  of these priors and investigated the aspect of model selection consistency under each prior in the normal linear regression model. Under the M-closed scenario, consistency is proved theoretically, while under the M-open case, consistency is investigated via simulation, using three different model misspecification scenarios.   

The main findings can be summarized as follows. 
In the Gaussian case, 
the DR-PEP prior coincides with the original conditional PEP prior of \cite{Fouskakis_Ntzoufras_2016_jcgs}, thus, sharing the same M-closed consistency and parsimony properties. 
On the other hand, the predictive distribution of the imaginary data used in the CR-PEP set-up, results in a PEP prior form which is less dispersed and, therefore, also less parsimonious than the DR-PEP prior. Nevertheless, the CR-PEP prior also leads to an M-closed consistent variable selection procedure. In addition, both priors have larger variances than the unit-information $g$-prior, which implies that they will support more parsimonious models than the $g$-prior. The DR-PEP prior in particular seems to be more suitable for large-$p$ problems, as its variance ratio over the $g$-prior increases as the number of predictor variables becomes larger.    

Generally, the resulting PEP priors can be viewed as extensions of the $g$-prior by considering imaginary, instead of fixed, data coming from a ``suitable" predictive distribution.
Specifically, they have an extra hierarchical level to the specification of the prior distribution
that has an effect on both the prior mean and the prior variance, through the variability of
the imaginary data. For more details see \cite{Fouskakis_Ntzoufras_2016_jcgs}.

Concerning the desiderata formulated in the important work of \cite{Bayarri_etal_2012}, the two new versions of PEP prior satisfy the basic criterion ($C1$) of propriety, the criterion of model consistency ($C2$), under the M-closed normal linear regression setup (see Section \ref{cons}), and the predictive matching criterion ($C5$), under the general GLM setup \citep{Fouskakis_Ntzoufras_2016_jcgs}. On the other hand both versions fail on the information consistency criterion ($C4$); see \cite{Fouskakis_Ntzoufras_2016_jcgs}. The latter issue can be resolved through the hyper-$\delta$  PEP prior extension introduced in \cite{Fouskakis_etal_2018_glm}. Therefore, an additional interesting direction of future research is to examine the properties of those hyper-$\delta$ versions of PEP priors, under the normal linear regression setup.  

Regarding the simulation experiments with the three M-open scenarios of model misspecification that we considered, the empirical findings suggest that PEP priors seem to be quite robust as the results (posterior inclusion probabilities, posterior probability of true model) we obtained were very similar to those under the M-closed case. Notably, that was also the case for the competing methods ($g$, hyper-$g$ and hyper-$g/n$ prior). Studying analytically the theoretical properties of PEP priors under model misspecification is a further interesting future research direction. 

Under the M-open case, using a proper inverse-gamma prior on $\sigma^2$, instead of the (improper) reference prior, would potentially be more effective in controlling the tails of the posterior of $\sigma^2$ in cases where the true generating mechanism has heavier tails and thus inflates the variance. In this work we preferred the standard (objective) choice of the reference prior because (a) the main aim was not estimation under model misspecification and (b) we focused on comparisons with other standard objective approaches (the $g$-prior and its extensions) that also use the reference prior.
Generally, the
reference prior is considered as a standard choice in the objective Bayesian variable selection literature (see for example
in \cite{liang_etal_2008}) since the error variance is treated as a common parameter under estimation.

Extension of the PEP prior methodology depends upon the specification of the power-likelihood. 
For Gaussian models, the power-likelihood can be written in a straightforward manner due to the property of the normal distribution which results again to a normal distribution with variance inflated  by a factor of $\delta$ when it is raised to the power of $\delta$.  
Hence, the PEP prior can be specified for any model that has in its core formulation the normal distribution; for example, Gaussian mixture models. For models with other sampling distributions this property is lost, as the corresponding power-likelihood
can be of non-standard form. The PEP variations discussed here, initially introduced in  \cite{Fouskakis_etal_2018_glm} can be
applied to generalized linear models and possibly models that have more complicated structures.
Efficient computation
in large spaces still remains an issue for future research for such cases. A possible solution can be found by approximating
PEP priors by simpler normal distributions as the ones proposed by \cite{Rodriguez_2015}. Of course, the BIC-consistency of
the PEP-based marginal likelihood, should need to be explored in model-specific context. Nevertheless, the general result
of \cite{schwarz_78} is still valid. 
Moreover, the unit-information interpretation of the PEP priors makes us intuitively believe that the results of \cite{kass_wasserman_95} can be extended also for PEP priors.

Implementation of the PEP methodology in high dimensional problems can be achieved using standard model search
tools, because the marginal likelihood under the proposed priors is readily available. Therefore, when full enumeration is computationally infeasible, model space exploration may be carried out by using the $MC^3$ algorithm \citep{madigan_york_95}, 
the strategies proposed in \cite{Johnson_Rossell_2012},  
the adaptive sampling scheme of \cite{clyde_etal_2011} or 
any other adaptive model search approach; see for example in \cite{Ji_Schmidler_2013}. 

Finally, it is worth nothing that the simulation-based findings presented here do not necessarily provide evidence for
cases of high–dimensional problems. According to a referee comment, robustness of posterior inclusion probabilities is
expected in low dimensions, since the covariate effect detection rate depends on model specification as well
as model dimensionality \citep{Rossell_Rubio2018}. In this work we considered misspecification of the distribution of the
residual errors. Misspecification of the functional form of the covariates is a further interesting future research direction.


\section*{Acknowledgements/Funding}
We wish to thank the Associate Editor and two
referees for comments that greatly strengthened the paper.\\

\noindent  This research was funded by the Research Centre of the Athens University of Economics
and Business (Funding program Action 2 for the support of basic research).

\newpage
\appendix

\section*{Appendix}


\section{Derivation of Equation \ref{CRvolume}}
\label{appendix_CRvolume}

The determinant of the CR-PEP prior covariance matrix is
\begin{equation}
\big|\VCR\big|=\delta^{d_{\vg}}\big|w^{-1}\XT\XG-\XT(\delta\LamdaOCR+w\HL)^{-1}\XG\big|^{-1}.
\label{a41}
\end{equation}
Based on the matrix determinant Lemma \citep[p.416]{harville_97}, which states that  $|\mathbf{A}+\mathbf{CBD}^T|=|\mathbf{A}||\mathbf{B}||\mathbf{B}^{-1}+\mathbf{D}^T\mathbf{A}^{-1}\mathbf{C}|$ for any square invertible matrices $\mathbf A$ and $\mathbf B$, we can write \eqref{a41} as
\begin{eqnarray}
\big|\VCR\big| &=& \delta^{d_{\vg}}\Big(
|w^{-1}\XT\XG|
|-(\delta\LamdaOCR+w\HL)|^{-1} \times\nonumber\\
&& \times
|-(\delta\LamdaOCR+w\HL)+w\XG(\XT\XG)\XT|\Big)^{-1} \nonumber \\
&=& \delta^{d_{\vg}}w^{d_{\vg}}|\XT\XG|^{-1}|\delta\LamdaOCR|^{-1}|(\delta\LamdaOCR+w\HL)|
\label{a42}.
\end{eqnarray}
Using repeatedly the matrix determinant Lemma on the last term of \eqref{a42} yields 
\begin{eqnarray}
|\delta \LamdaOCR + w \HL| & = & |\delta \LamdaOCR| |\XT\XG|^{-1} \left|\XT\XG+\frac{w}{\delta}\XT\invLamdaOCR\XG\right| \nonumber \\
& = &
|\delta \LamdaOCR| |\XT\XG|^{-1} \left|\XT\XG+\frac{w}{\delta}\XT\left(\I+g_0\XO({\XO}^T\XO)^{-1}{\XO}^T\right)\XG\right| \nonumber \\
& = &
|\delta \LamdaOCR| |\XT\XG|^{-1} \left|\XT\XG+\frac{w}{\delta}\XT\XG+\frac{wg_0}{\delta}\XT\XO({\XO}^T\XO)^{-1}{\XO}^T\XG\right| \nonumber \\
& = &
|\delta \LamdaOCR| |\XT\XG|^{-1} \left|\frac{w+\delta}{\delta}\XT\XG+\frac{wg_0}{\delta}\XT\XO({\XO}^T\XO)^{-1}{\XO}^T\XG\right| \nonumber \\
& = &
|\delta \LamdaOCR| |\XT\XG|^{-1} \left(\frac{w+\delta}{\delta}\right)^{d_{\vg}}|\XT\XG|
|{\XO}^T\XO|^{-1} \times \nonumber \\
& & \times \left|{\XO}^T\XO+\frac{wg_0}{\delta}
{\XO}^T\XG\left(\frac{w+\delta}{\delta}\XT\XG\right)^{-1}\XT\XO \right| \nonumber \\
& = &
\left(\frac{w+\delta}{\delta}\right)^{d_{\vg}}|\delta \LamdaOCR| 
|{\XO}^T\XO|^{-1}\times\nonumber\\
&&\times \left|{\XO}^T\XO+\frac{wg_0}{w+\delta}
{\XO}^T\XG\left(\XT\XG\right)^{-1}\XT\XO \right| \label{a43}  \\
& = &
\left(\frac{w+\delta}{\delta}\right)^{d_{\vg}}|\delta \LamdaOCR| 
|{\XO}^T\XO|^{-1}\left|{\XO}^T\XO+\frac{wg_0}{w+\delta}
{\XO}^T\XO\right| \nonumber \\
& = &
\left(\frac{w+\delta}{\delta}\right)^{d_{\vg}}|\delta \LamdaOCR| 
|{\XO}^T\XO|^{-1}\left(\frac{w+\delta+ wg_0}{w+\delta}\right)^{d_0}|{\XO}^T\XO| \nonumber\\
& = &
\left({w+\delta}\right)^{d_{\vg}-d_0}\delta^{-d_{\vg}} \left(w+\delta+ wg_0\right)^{d_0}|\delta \LamdaOCR| 
\label{a44} 
\end{eqnarray}
Note that the transition from \eqref{a43} to the following equation is due to the fact that ${\XO}^T\XG\left(\XT\XG\right)^{-1}\XT\XO={\XO}^T\HL\XO={\XO}^T\XO$, since ${\XO}^T\HL={\XO}^T$ for any sub-matrix $\XO$ of $\XG$.
From \eqref{a42} and \eqref{a44} we have that
\begin{equation}
|\VCR| = w^{d_{\vg}}\left({w+\delta}\right)^{d_{\vg}-d_0}\left(w+\delta+ wg_0\right)^{d_0}|\XT\XG|^{-1}. 
\end{equation}

\section{Derivation of Equation \ref{RSS_CR}}
\label{appendix_RSS_CR} 

\vspace{0.01em}
\begin{align}
&{\by}^T(\I+\XG\VCR\XT)^{-1}\by \nonumber \\
& = \by^Y\Big(\IotaN-\XG\big(\invVCR+\XT\XG\big)^{-1}\XT\Big)\by \nonumber\\
& =  {\by}^T\by- \by^T\XG\Bigg(
\delta^{-1}\Big[
{\X}^{T}\Big(
w^{-1}\I-\big(\delta\LamdaOCR+w\HL\big)^{-1}
\Big)\X\Big]
+\XT\XG\Bigg)^{-1}\XT\by \nonumber\\
& =  {\by}^T\by- \delta{\by}^T\XG\Bigg(
w^{-1}\XT\XG- 
\XT\Big(\delta \LamdaOCR + w\HL\Big)^{-1}\XG+  \delta\XT\XG
\Bigg)^{-1}\XT\by \nonumber \\
& = {\by}^T\by- \delta{\by}^T\XG\Bigg(
\frac{1+\delta w}{w}\XT\XG-\XT\Big(\delta \LamdaOCR + w\HL\Big)^{-1}\XG\Bigg)^{-1}\XT\by \nonumber \\
& = {\by}^T\by- 
\delta{\by}^T\XG\Bigg(
\frac{1+\delta w}{w}\XT\XG-\XT\Big(\delta \Big(\I-\frac{g_0}{g_0+1}\mathbf{H}_0\Big) + w\HL\Big)^{-1}\XG\Bigg)^{-1}\XT\by \nonumber \\
& ={\by}^T\by- 
\frac{w\delta}{1+w\delta}{\by}^T\XG\Bigg(
\XT\XG-\frac{w}{1+w\delta}\XT\Big(\delta \Big(\I-\frac{g_0}{g_0+1}\mathbf{H}_0\Big) + w\HL\Big)^{-1}\XG\Bigg)^{-1}\XT\by \nonumber \\
& ={\by}^T\by- 
\frac{w\delta}{1+w\delta}{\by}^T\XG\Bigg(
\XT\XG-\frac{w}{(1+w\delta)\delta}\XT\Big(\I-\frac{g_0}{g_0+1}\mathbf{H}_0 + \frac{w}{\delta}\HL\Big)^{-1}\XG\Bigg)^{-1}\XT\by, 
\end{align}
where $\HL=\XG(\XT\XG)^{-1}\XT$ and $\mathbf{H}_0=\XO(\XO^T\XO)^{-1}\XO^T=\IotaN(\IotaN^T\IotaN)^{-1}\IotaN^T=n^{-1}\IotaN\IotaN^T$. For the derivation of the first expression, see Woodbury’s matrix identity \citep[p. 423--426]{harville_97}. For large values of $\delta$ and $g_0>>\delta$ we have approximately $w\approx 1$, $\frac{w\delta}{1+w\delta}\approx1$ and $\frac{w}{(1+w\delta)\delta}\approx0$, which yields the approximation  
\begin{equation}
{\by}^T(\I+\XG\VCR\XT)^{-1}\by \approx {\by}^T\by-{\by}^T\XG\left(
\XT\XG
\right)^{-1}\XT\by 
\equiv \mathrm{RSS}_{\vg}. 
\end{equation}

\clearpage
\section{Additional results from Section \ref{S1}}
\label{Sim1} 
\renewcommand{\thefigure}{C.\arabic{figure}}
\renewcommand{\theHfigure}{C.\arabic{figure}}  
\setcounter{figure}{0} 

\begin{figure}[H]
	\centering{}
	\includegraphics[scale=0.67]{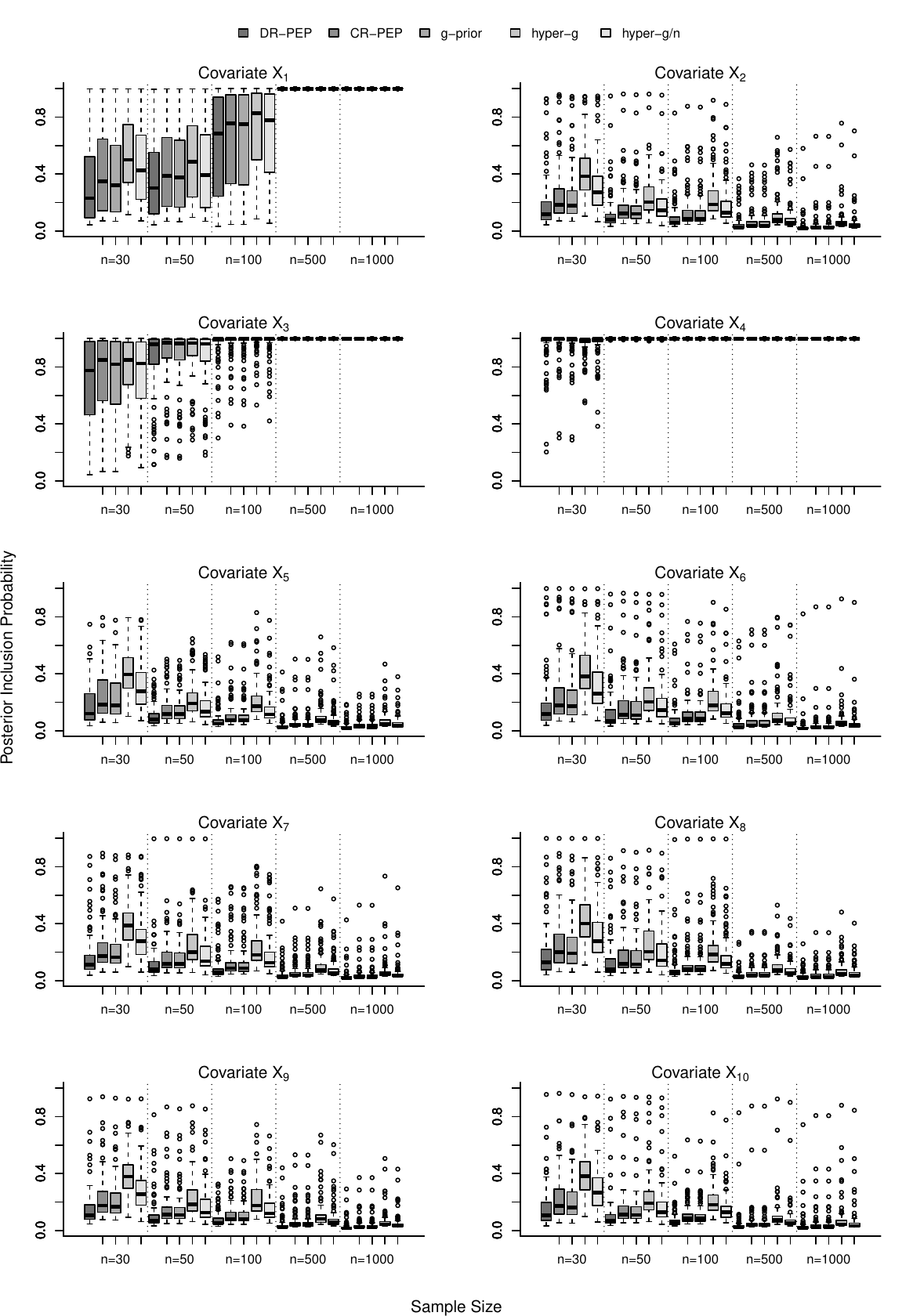}
	\caption{Posterior  inclusion probabilities of each covariate for Scenario 1 (independent covariates) under Laplace errors (M-open case).}
	\label{inc_probs_ex1_laplace}
\end{figure}

\begin{figure}[H]
	\centering{}
	\includegraphics[scale=0.67]{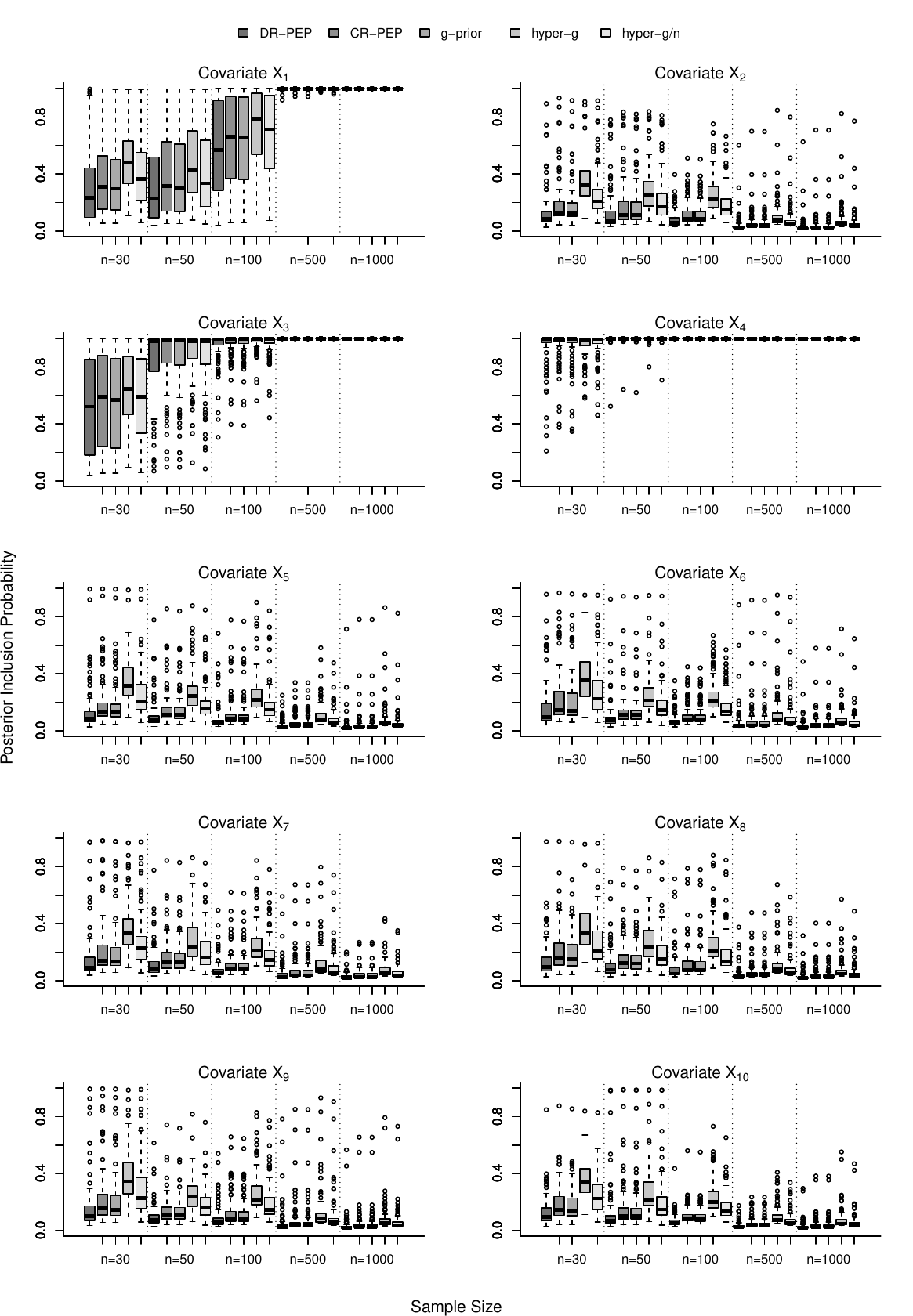}
	\caption{Posterior  inclusion probabilities of each covariate for Scenario 1 (independent covariates) under asymmetric normal errors (M-open case).}
	\label{inc_probs_ex1_a_normal}
\end{figure}

\begin{figure}[H]
	\centering{}
	\includegraphics[scale=0.67]{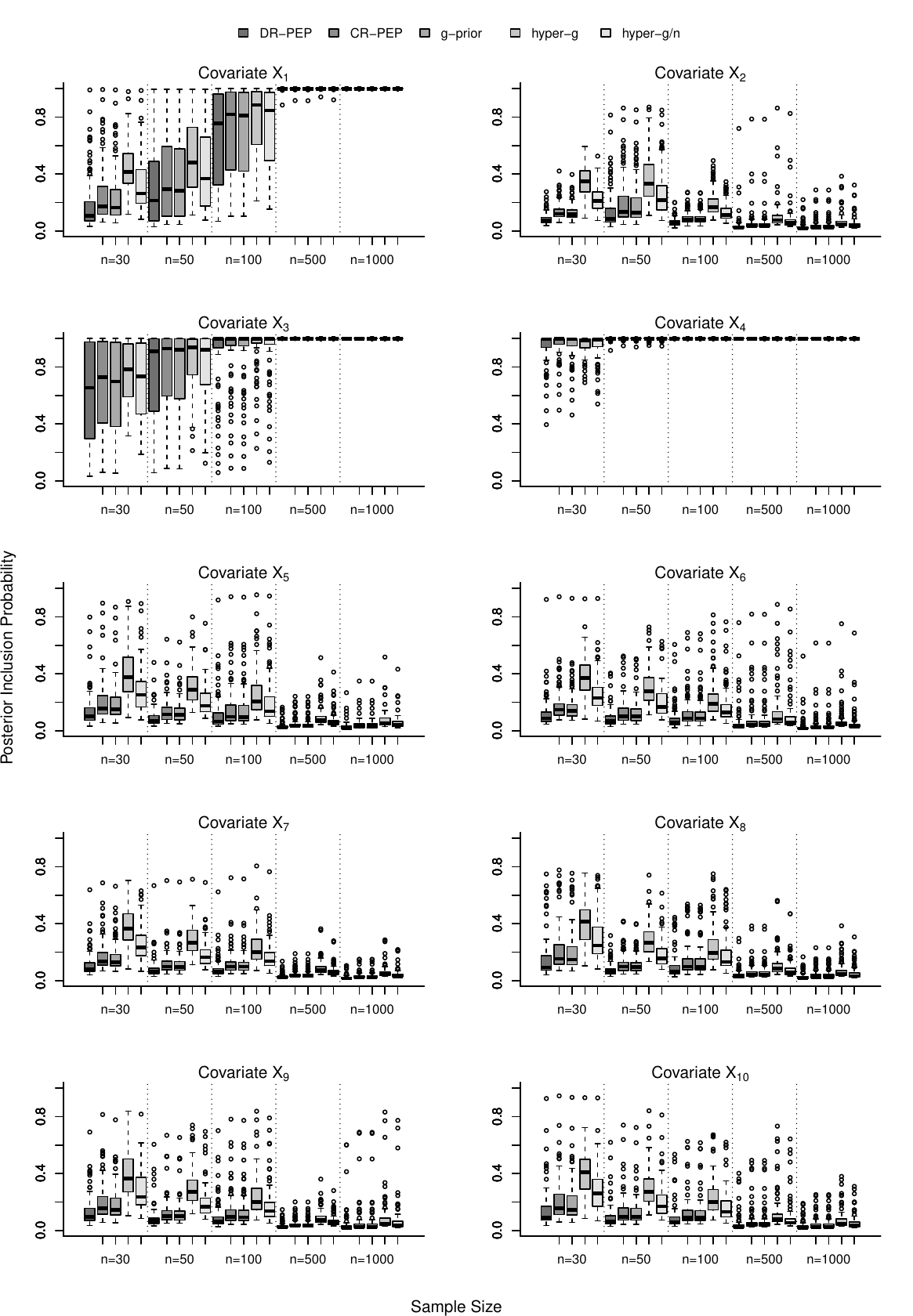}
	\caption{Posterior  inclusion probabilities of each covariate for Scenario 1 (independent covariates)  heteroscedastic normal errors (M-open case).}
	\label{inc_probs_ex1_h_normal}
\end{figure}

\newpage
\section{Additional results from Section \ref{S2}}
\label{Sim2} 
\renewcommand{\thefigure}{D.\arabic{figure}}
\renewcommand{\theHfigure}{D.\arabic{figure}}  
\setcounter{figure}{0} 

\begin{figure}[H]
	\centering{}
	\includegraphics[scale=0.67]{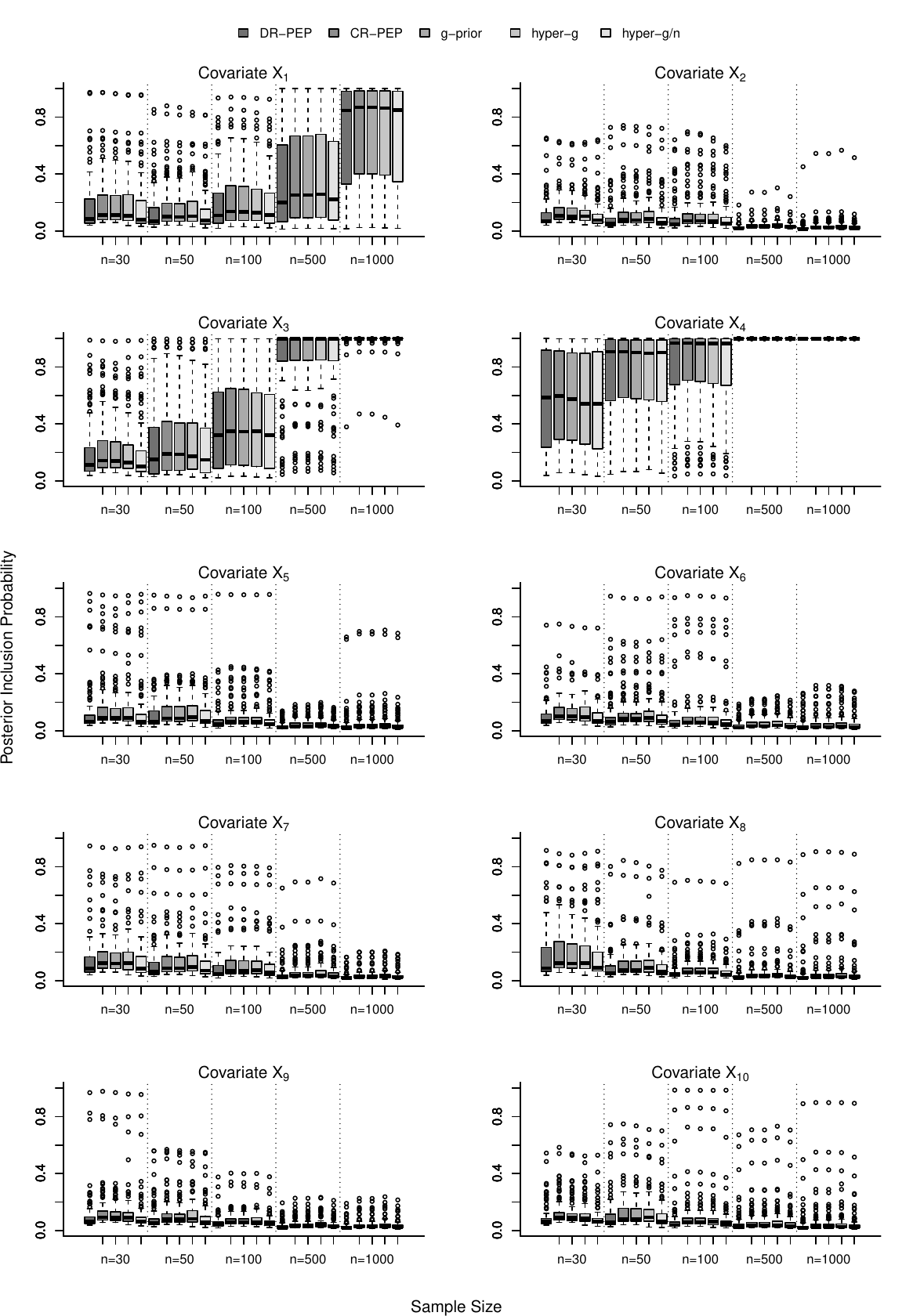}
	\caption{Posterior  inclusion probabilities of each covariate for Scenario 2 (highly-correlated covariates) under Laplace errors (M-open case).}
	\label{inc_probs_ex2_laplace}
\end{figure}

\begin{figure}[H]
	\centering{}
	\includegraphics[scale=0.67]{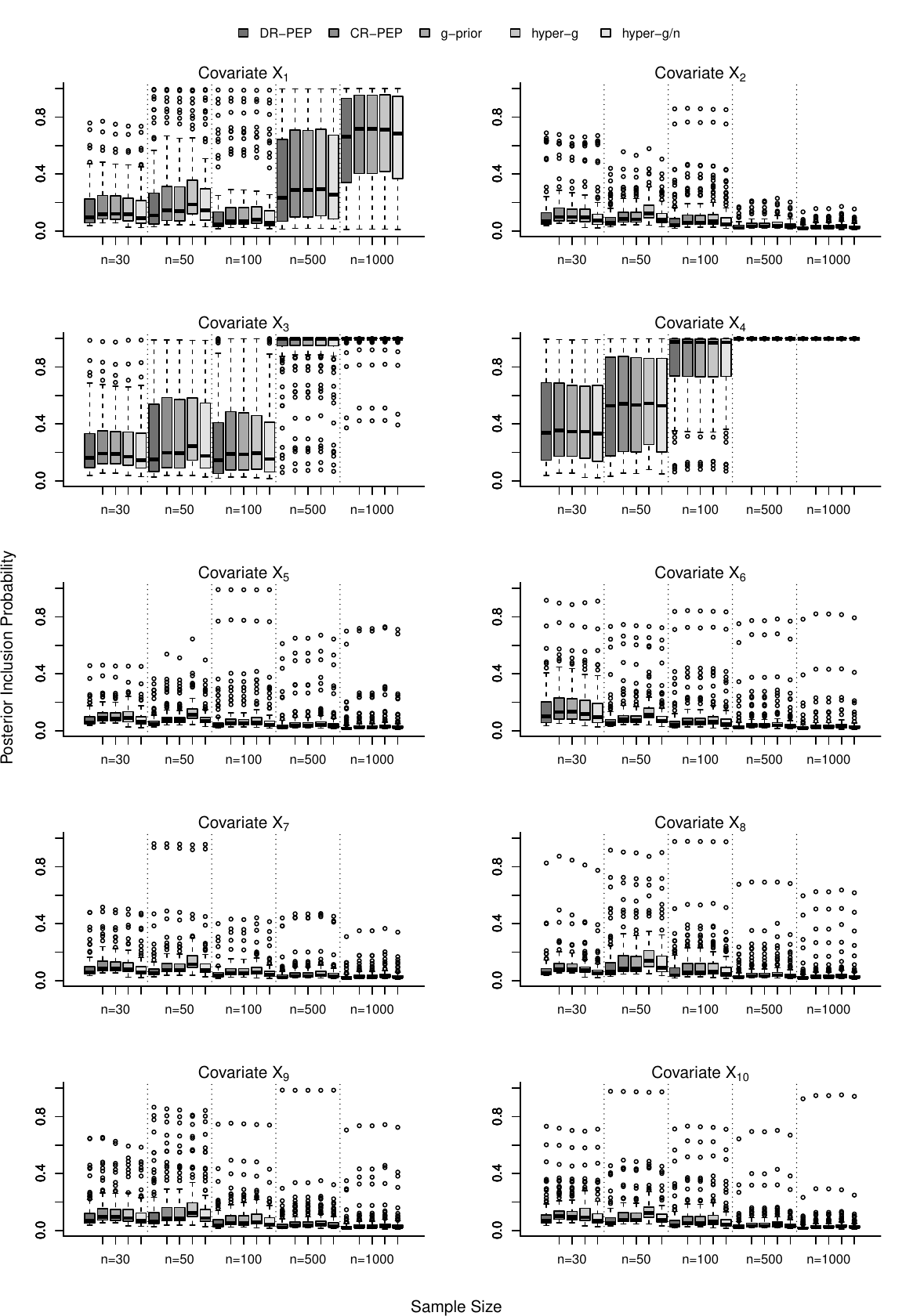}
	\caption{Posterior  inclusion probabilities of each covariate for Scenario 2 (highly-correlated covariates) under asymmetric normal errors (M-open case).}
	\label{inc_probs_ex2_a_normal}
\end{figure}

\begin{figure}[H]
	\centering{}
	\includegraphics[scale=0.67]{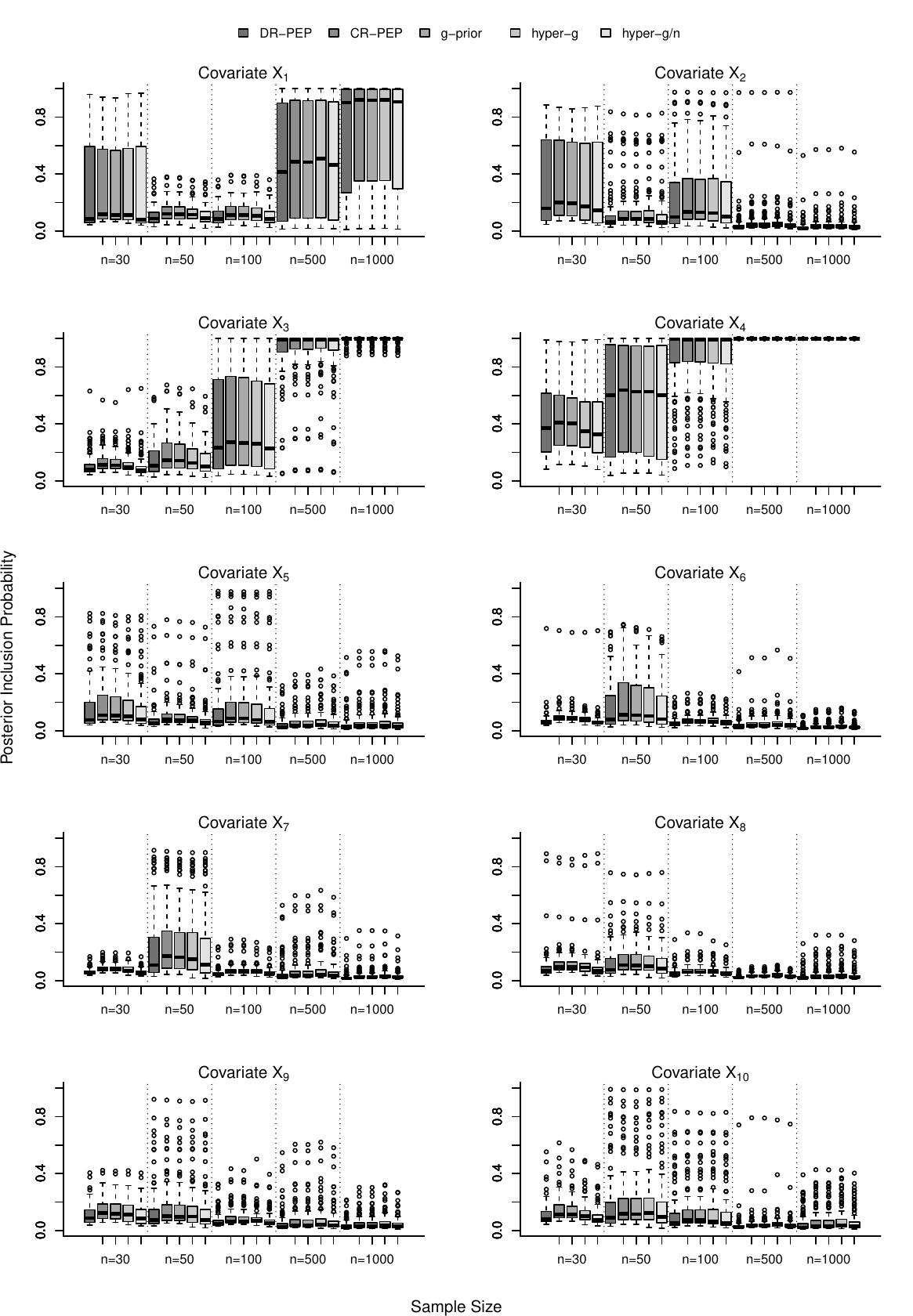}
	\caption{Posterior  inclusion probabilities of each covariate for Scenario 2 (highly-correlated covariates)  heteroscedastic normal errors (M-open case).}
	\label{inc_probs_ex2_h_normal}
\end{figure}
 
 \newpage
 \section{Additional results from Section \ref{S_p}}
 \label{AppSim3} 
 \renewcommand{\thefigure}{E.\arabic{figure}}
 \renewcommand{\theHfigure}{E.\arabic{figure}}  
 \setcounter{figure}{0}

\begin{figure}[H]
	\centering{}
	\includegraphics[scale=0.43, angle=-90]{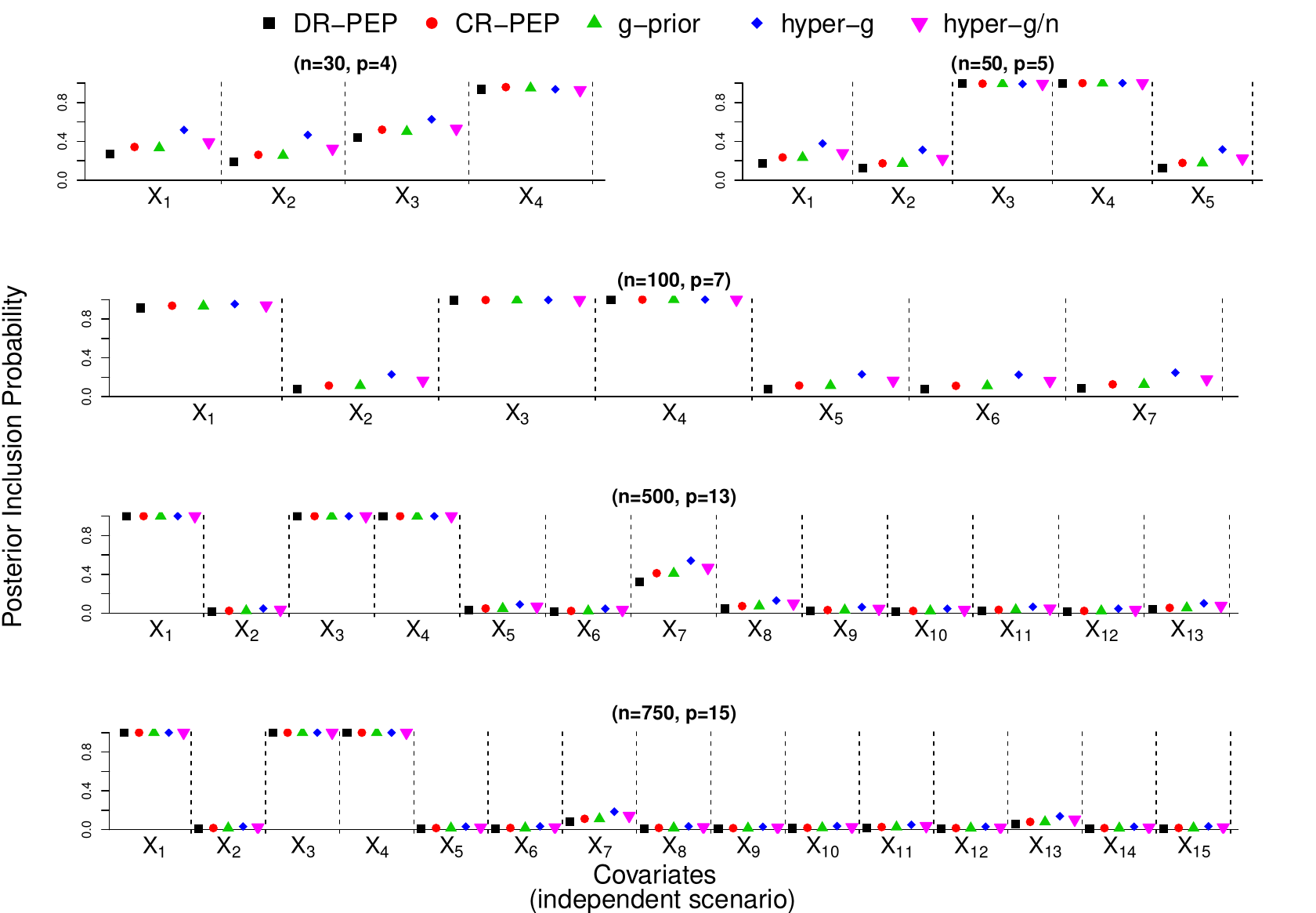}
	\caption{Posterior inclusion probabilities of each covariate for different sample sizes and number of covariates for the independence scenario.}
	\label{inc_probs3}
\end{figure}
\begin{figure}[H]
	\centering{}
	\includegraphics[scale=0.43, angle=-90]{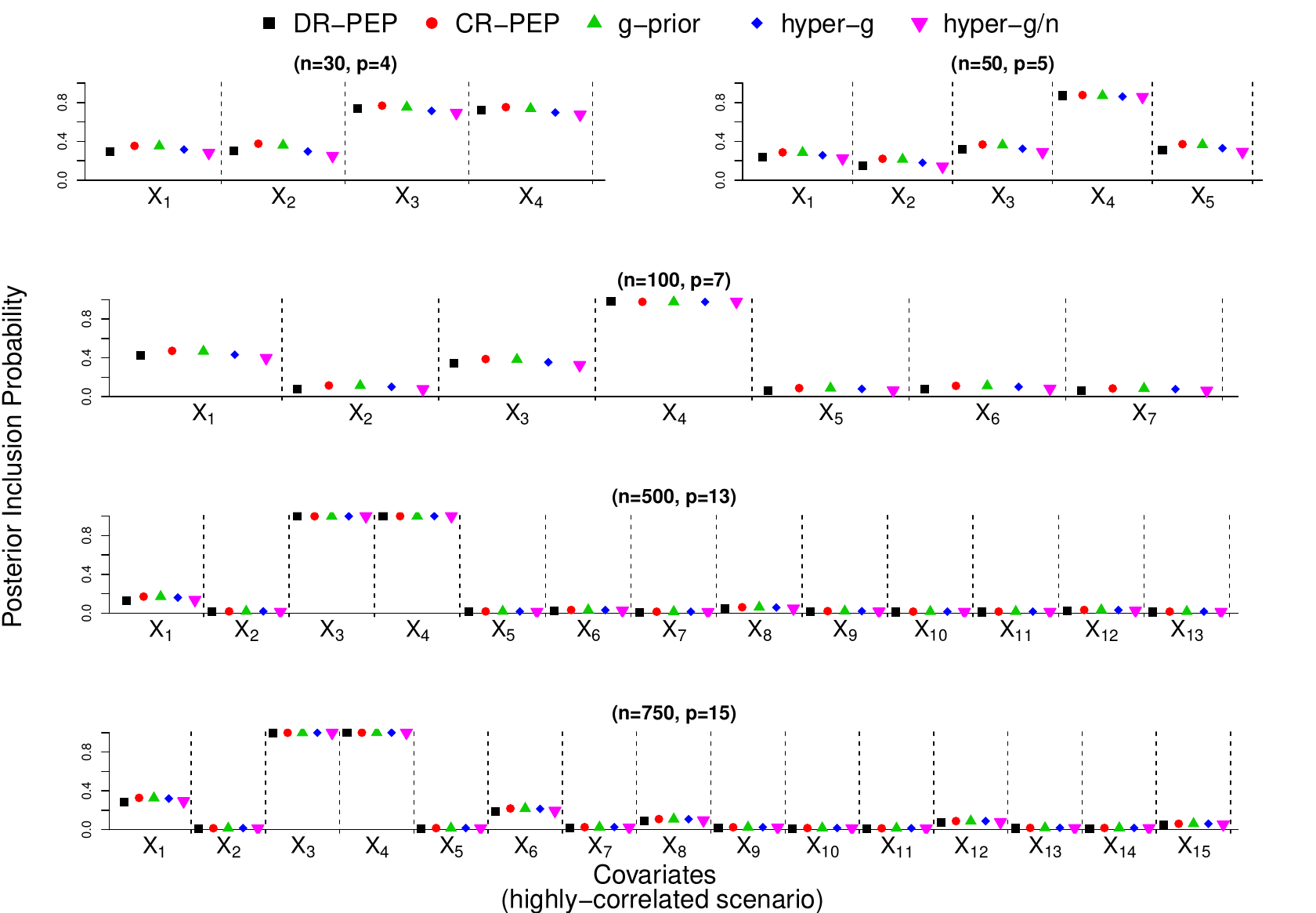}
	\caption{Posterior  inclusion probabilities of each covariate for different sample sizes and number of covariates for the high-correlation scenario.}
	\label{inc_probs4}
\end{figure}

\newpage
\bibliographystyle{agsm}

\bibliography{biblio2016}

\end{document}